\documentclass[
 reprint, superscriptaddress,
 amsmath,amssymb,
 aps,
prb,
]{revtex4-2}

\usepackage{graphicx}
\usepackage{dcolumn}
\usepackage{bm}
\usepackage[colorlinks=true,citecolor=blue,linkcolor=blue]{hyperref}

\usepackage{booktabs}

\begin{document}

\preprint{APS/123-QED}

\title{Origin of magic angles in twisted bilayer graphene: The magic ring}

\author{Wei-Chen Wang}
\thanks{These authors contributed equally to this work.}
\affiliation{Institute of Atomic and Molecular Sciences, Academia Sinica, Taipei 106319, Taiwan}
\affiliation{Department of Physics, National Taiwan University, Taipei 106319, Taiwan}
\author{Feng-Wu Chen}
\thanks{These authors contributed equally to this work.}
\affiliation{Institute of Atomic and Molecular Sciences, Academia Sinica, Taipei 106319, Taiwan}
\author{Kuan-Sen Lin}
\affiliation{Institute of Atomic and Molecular Sciences, Academia Sinica, Taipei 106319, Taiwan}
\affiliation{Department of Physics, National Taiwan University, Taipei 106319, Taiwan}
\author{Justin T. Hou}
\affiliation{Institute of Atomic and Molecular Sciences, Academia Sinica, Taipei 106319, Taiwan}
\affiliation{Department of Physics, National Taiwan University, Taipei 106319, Taiwan}
\author{Ho-Chun Lin}
\affiliation{Institute of Atomic and Molecular Sciences, Academia Sinica, Taipei 106319, Taiwan}
\affiliation{Department of Physics, National Taiwan University, Taipei 106319, Taiwan}
\author{Mei-Yin Chou}
\email[Corresponding author. Email: ]{mychou6@gate.sinica.edu.tw}
\affiliation{Institute of Atomic and Molecular Sciences, Academia Sinica, Taipei 106319, Taiwan}
\affiliation{Department of Physics, National Taiwan University, Taipei 106319, Taiwan}


\begin{abstract}
The unexpected discovery of superconductivity and strong electron correlation in twisted bilayer graphene (TBG), a system containing only \textit{sp} electrons, is considered as one of the most intriguing developments in two-dimensional materials in recent years. The key feature is the emergent flat energy bands near the Fermi level, a favorable condition for novel many-body phases, at the so-called ``magic angles". The physical origin of these interesting flat bands has been elusive to date, hindering the construction of an effective theory for the unconventional electron correlation. In this work, we have identified the importance of charge accumulation in the AA region of the moir\'{e} supercell and the most critical role of the Fermi ring in AA-stacked bilayer graphene. We show that 
the magic angles can be predicted by the moir\'{e} periodicity determined by the size of this Fermi ring.
The resonant criterion in momentum space makes it possible to coherently combine states on the Fermi ring through scattering by the moir\'{e} potential, leading to flat bands near the Fermi level. We thus establish the physical origin of the magic angles in TBG and identify the characteristics of one-particle states associated with the flat bands for further many-body investigations.
\end{abstract}

\maketitle


\section{\label{sec:intro}Introduction}
Two graphene sheets with a {\it small} rotation angle $\theta$ between them create a special system of twisted bilayer graphene (TBG), in which a moir\'{e} pattern emerges with a spatial periodicity inversely proportional to the twist angle \cite{PhysRevB.81.165105,PhysRevB.86.155449}. It has been shown that the variation of the twist angle in moir\'{e} two-dimensional (2D) materials could modify the electronic properties, giving rise to an interesting research area of twistronics \cite{PhysRevB.95.075420}. Previous calculations on TBG using different theoretical methods including the low-energy continuum model \cite{PhysRevLett.99.256802,PhysRevB.86.155449,Bistritzer12233}, the tight-binding method \cite{PhysRevB.81.165105,PhysRevB.82.121407,tramblydelaissardiere2010,tramblydelaissardiere2012}, and density functional theory (DFT) \cite{uchida2014} all had similar findings: as $\theta$ decreases, the Fermi velocity of the linear bands also becomes progressively smaller. This results from an enhanced interlayer hybridization between the two rotated Dirac cones of different layers as they get closer in momentum space when $\theta$ gets smaller. These calculations also revealed that as the bandwidth of the low-lying bands is reduced at small $\theta$ angles, the corresponding electron charge becomes localized to the AA-stacked region in the moir\'{e} supercell \cite{PhysRevB.86.155449,tramblydelaissardiere2010,uchida2014}. One would anticipate that the Fermi velocity should decrease smoothly as $\theta$ approaches zero; however, it was found in these calculations that at certain so-called ``magic angles" the velocity deviates away from the expected monotonic curve and drops to almost zero \cite{Bistritzer12233,PhysRevB.86.155449,tramblydelaissardiere2012}, with the first of these magic angles being around $1.1^{\circ}$ . The resulting unusually flat bands thus have a large ratio between the Coulomb and kinetic energies and favor the formation of possible many-body phases.

Experimentally, with the ``tear-and-stack" technique \cite{cao2016,kim2016,kim2017,frisenda2018, ribeiro-palau2018}, researchers have been able to control the rotation angle of TBG precisely. It was first discovered by Cao \textsl{et al.} \cite{cao2018a} that around the first magic angle there exist insulating gaps when the doping level is tuned to integer fillings. Moreover, superconducting states appear \cite{cao2018b} in the intermediate filling regions. These surprising reports stimulated further transport measurements for detailed studies on correlated insulating states at several integer fillings \cite{lu2019}, superconductivity domes in a wide, continuous range of doping \cite{lu2019,saito2020,stepanov2020,cao2021a,liu2021a}, ferromagnetism \cite{lu2019,sharpe2019}, and the integer quantum anomalous Hall effect \cite{serlin2020} at certain odd fillings. The existence of flat bands in TBG was confirmed by scanning tunneling spectroscopy (STS) studies \cite{Guohong2010,PhysRevLett.109.196802,Kerelsky2019} that revealed two pronounced van Hove singularities very close to each other in the density of states (DOS) near the first magic angle . Using angle-resolved photoemission spectroscopy with nanoscale resolution (nano-ARPES), energy bands with little dispersion were directly observed in momentum space at $\theta \approx 0.96^{\circ}$ \cite{utama2021} and $1.34^{\circ}$ \cite{Lisi2021}. The charge localization in the AA region at small $\theta$ was also confirmed using scanning tunneling microscope (STM) \cite{kim2017,Guohong2010,PhysRevLett.109.196802,Kerelsky2019}. Furthermore, both transport and STS measurements in the presence of magnetic field have recently identified correlated gaps that account for both integer \cite{nuckolls2020,choi2021a,wu2021,saito2021a,das2021a,park2021a,pierce2021} and fractional \cite{wu2021,xie2021b} Chern insulating states, as a result of strong correlation in combination with the flat band topology.

It is unprecedented that a carbon material with only $sp$ electrons can exhibit such rich phenomena of electron correlation driven by interaction. Thus, magic-angle twisted bilayer graphene (MATBG) and related systems have presented special challenges in our understanding of two-dimensional physics. In order to explain these exotic properties observed, various theoretical investigations have been reported. Most of these studies \cite{yuan2018,kang2018,koshino2018,po2018b} relied on building a minimum model/basis set consisting of, for example, Wannier orbitals, which reproduces the single-particle flat band dispersion while satisfying the original symmetry and topology properties as much as possible. Many-body interactions were then included in the framework of (generalized) Hubbard model to shed light on the nature of many-body states observed experimentally. Some other studies \cite{rademaker2018,xie2020a,PhysRevLett.129.047601} pursued alternative approaches to avoid possible bias toward the basis set.
Nevertheless, no convincing theory that has been proposed to date can fully explain the interesting and intriguing experimental findings in MATBG yet.

The existence of multiple magic angles with extremely flat bands was found in many one-particle calculations. Yet their origin and why a series of flat bands occur at these specific magic angles are still not understood until now. Building an effective many-body theory for electron correlation and superconductivity requires, as the starting point, a good knowledge of the characteristics of one-particle electronic states at the magic angles. Therefore, in this theoretical work we aim to uncover the unique features in the electronic properties of TBG.  Since the system is too large for full-scale DFT calculations, we mostly used the tight-binding method with newly developed parameters for accurate angle-dependent interlayer interactions \cite{PhysRevB.93.235153, PhysRevB.98.085144}. First, we provide an explanation using DFT calculations for the reason why electrons near the Fermi level become localized in the AA region of TBG when the twist angle $\theta$ gets small. As a result, the local electronic structure of AA-stacked bilayer graphene (AABLG) with a characteristic Fermi ring becomes the key feature to be considered for generating a special effect at certain twist angles. Second, we demonstrate that when the size of this Fermi ring matches the reciprocal lattice vectors of the moir\'{e} superlattice at a set of angles, the Dirac points of TBG fall on the Fermi ring. Multiple states on the ring can then be coupled coherently by the moir\'{e} potential, leading to energy bands with extremely small dispersions. This matching condition in momentum space generates a series of discrete magic angles in TBG. Computational results on TBG with different interlayer coupling strengths when it is under uniaxial pressure are presented to illustrate our theory. This work provides the physical explanation on the origin of magic angles. The result also indicates that magic angles are special features in twisted graphene systems arising from their particular electronic structure.

\section{Results}
\subsection*{Electron accumulation in the AA region}

\begin{figure*}[ht]
\centering 
\includegraphics[width=.8\linewidth]{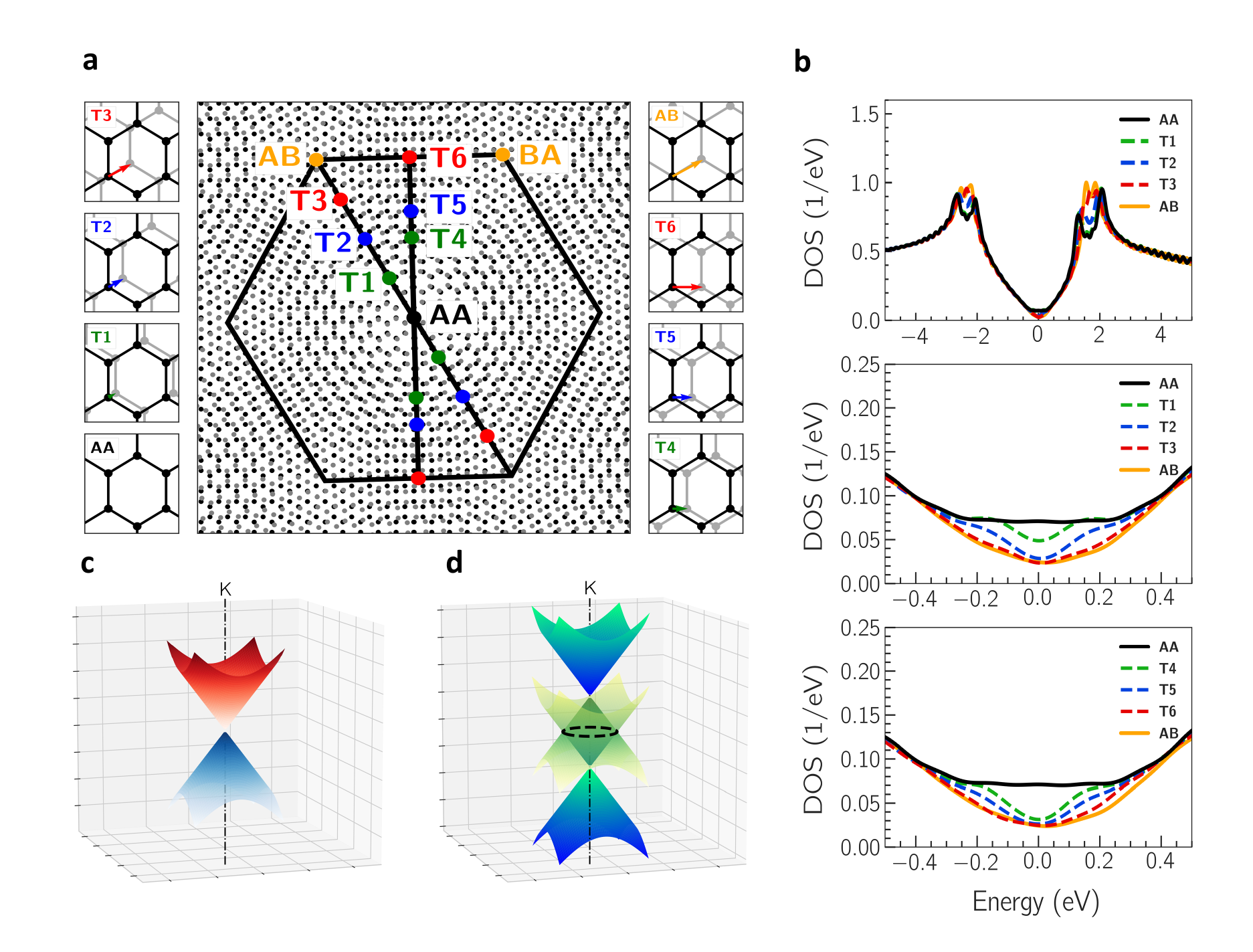}
\caption{$\mid$ {\bf Moir\'{e} effect on electronic density of states.} {\bf a}, Moir\'{e} pattern of TBG at $\theta=3.15^{\circ}$ with six representative local atomic configurations $\mathrm{T1}$ - $\mathrm{T6}$ in addition to AA and AB(BA). The black (gray) atoms are in the first (second) layer. The arrow in the insets shows how the second layer of AA-stacked graphene is shifted to generate a particular local stacking configuration. The thick hexagon marks the Wigner-Seitz cell of the moir\'{e} supercell. {\bf b}, Density of states (DOS) of representative bilayer stacking configurations in {\bf a}, obtained from density-functional-theory (DFT) calculations. The uppermost panel covers a larger energy range, while the lower two panels compare the DOS variations near the Fermi level along two paths ($\textrm{AA}\rightarrow\mathrm{T1}\rightarrow\mathrm{T2}\rightarrow \mathrm{T3}\rightarrow\textrm{AB}$ and $\textrm{AA}\rightarrow\mathrm{T4}\rightarrow\mathrm{T5}\rightarrow \mathrm{T6}$). It is noted that the AA region has significantly more low-energy states around the Fermi level. {\bf c}, Dirac cone in momentum space for monolayer graphene. {\bf d}, Two Dirac cones shifted in energy for AA-stacked bilayer graphene with the Fermi ring marked by a dashed circle.}
\label{fig:stacking}
\end{figure*}

In our tight-binding calculation, we have considered commensurate TBG configurations with moir\'{e} lattice vectors of $\mathbf{L_1} = m\,\mathbf{a_1}+(m+1)\,\mathbf{a_2}$ and $\mathbf{L_2} = -(m+1)\,\mathbf{a_1}+(2m+1)\,\mathbf{a_2}$, where $\mathbf{a_1}$ and $\mathbf{a_2}$ are lattice vectors for monolayer graphene, and $m$ is an integer \cite{PhysRevB.86.155449}.
The rotation starts from AABLG with the axis passing through a vertical pair of carbon atoms. We neglect the possible interlayer corrugation and intralayer relaxation in this work.
Figure \ref{fig:stacking}a shows the atomic arrangement of a typical moir\'{e} pattern in TBG with a rotation angle of $\theta= 3.15^{\circ}$ ($m=10$). Because of the relative rotation between two graphene layers, the local atomic arrangements exhibit different stacking patterns at different locations. Some noticeable ones, such as the AA-stacked and Bernal-stacked (AB or BA) regions, are labeled in Fig. \ref{fig:stacking}a, as well as six other representative stacking patterns (T1-T6) in between. Thus, the Bloch electrons will experience different lattice potentials in different stacking regions within a single moir\'{e} supercell, which is an important feature for this system. To understand the stacking effect, we have calculated the electronic DOS for these different lattice potentials with DFT, and the results are shown in Fig. \ref{fig:stacking}b. It turns out that the AA stacking gives significantly more low-energy states near the Fermi level, while the AB(BA) stacking has the fewest states in the same energy range. This can be understood by examining the band dispersion. Compared with monolayer graphene with one Dirac cone (Fig. \ref{fig:stacking}c), AABLG has two cones shifted vertically in energy (Fig. \ref{fig:stacking}d), giving rise to a sizable constant DOS between the two original band crossings. In contrast, the Bernal stacking gives roughly two low-energy parabolic bands joined at the Dirac point with a smaller DOS. A tight-binding model with only the nearest-neighbor interlayer interaction will give a ratio of 4:1 for the low-energy DOS in these two systems. As the size of the moir\'{e} superlattice increases with decreasing $\theta$, so is the size of a region with a particular stacking pattern. Therefore, this explains the observation that, at small $\theta$ values, the electrons near the Fermi level tend to accumulate around the AA region in the moir\'{e} cell, because more states are available to them there.



\begin{figure*}[ht]
\centering 
\includegraphics[width=.8\linewidth]{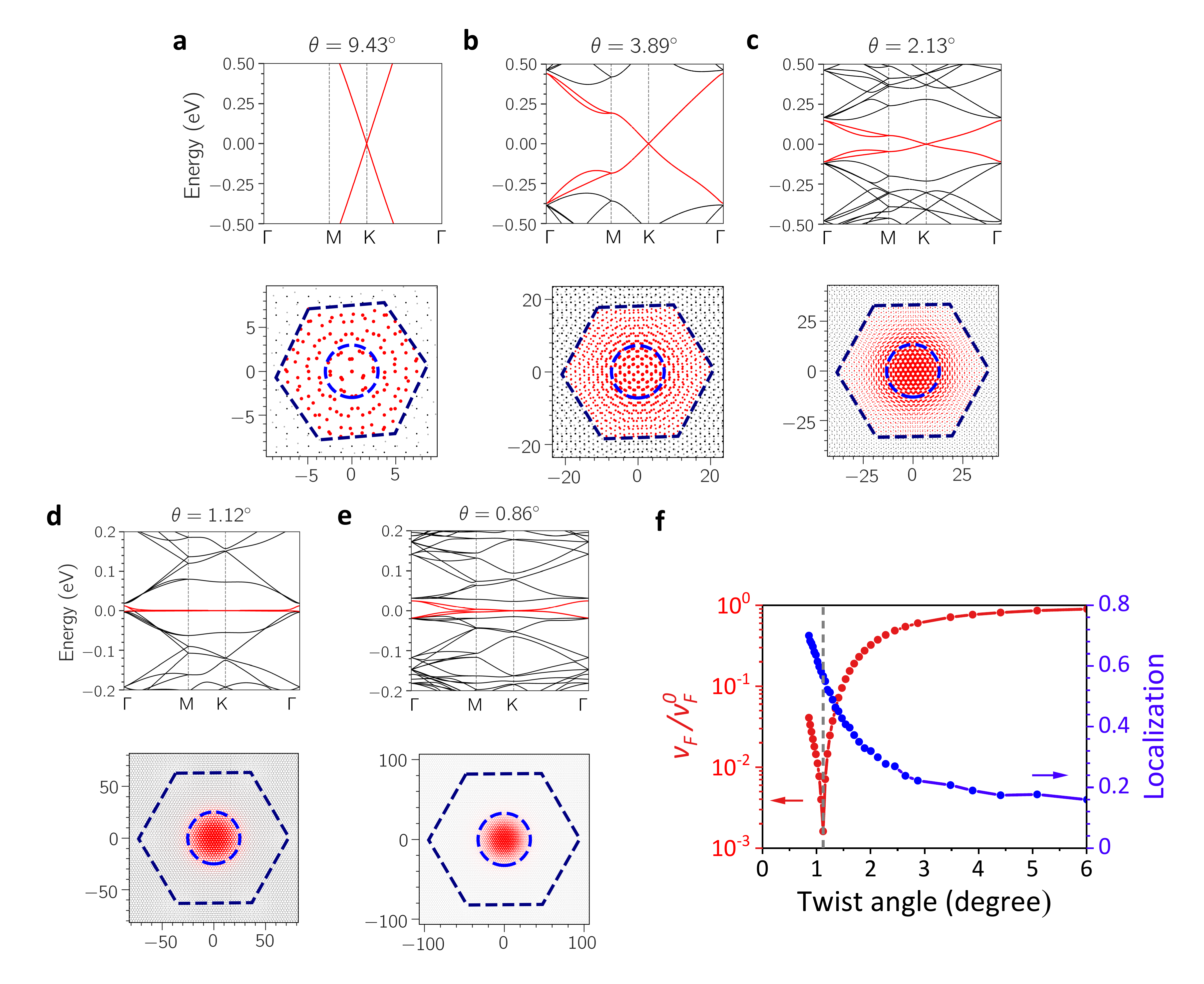}
\caption{$\mid$ {\bf Fermi velocity and localized electrons.} {\bf a-e}, Band structure (upper panel) of twisted bilayer graphene with angles of $9.43^\circ$, $3.89^\circ$, $2.13^\circ$, $1.12^\circ$ and $0.86^\circ$, respectively, calculated by the tight-binding method. The four bands near the Fermi level (energy zero) are marked in red. Note that the bands become flatter when the angle decreases, with a minimal Fermi velocity near $1.12^\circ$, but the slope at $0.86^\circ$ bounces back after passing through this magic angle. The lower panel shows the electronic probability distribution of the zero-energy states at $K$. The dark and light dots mark the atoms in layer 1 and 2, respectively, and the dashed dark hexagon represents the Wigner-Seitz cell with the real-space dimension labels in units of \AA. The size of the red dots represents the probability of finding the electrons at each atom.  {\bf f}, Calculated Fermi velocity (red curve) and electron localization (blue curve) as a function of the twist angle $\theta$. The Fermi velocity is normalized with respect to that in monolayer graphene $v_F^0$. The electron localization is measured by the total fraction of $K$ electrons within the the blue dashed circle centered at the AA region that has a radius of 20\% of the moir\'{e} lattice constant. The Fermi velocity has a sharp dip at $1.12^\circ$ near the magic angle. In contrast, the electron localization measure increases smoothly and monotonically as the twist angle decreases.}
\label{fig:velocity}
\end{figure*}

To illustrate the major change in the electronic structure as $\theta$ varies, we present in the upper panels of Fig. \ref{fig:velocity}a-e the band structure of TBG by tight-binding calculations for five different twist angles (9.43$^\circ$, 3.89$^\circ$, 2.13$^\circ$, 1.12$^\circ$ and 0.86$^\circ$). Note the change of the energy scale in these plots. The size of the moir\'{e} supercell is 1.5 nm, 3.6 nm, 6.6 nm, 13 nm, and 16 nm, respectively. The four bands in red near the Fermi energy gradually flatten as the twist angle decreases in Fig. \ref{fig:velocity}a-c, which is expected for an increased coupling between the Dirac cones of two layers. Then there is a sudden change near the first magic angle of 1.1$^\circ$ where the bands become extremely flat (Fig. \ref{fig:velocity}d). After passing through the first magic angle, the bands regain their slopes (Fig. \ref{fig:velocity}e). This trend can be seen clearly from Fig. \ref{fig:velocity}f in which the red curve shows the normalized Fermi velocity of TBG (with respect to the Fermi velocity $v^0_F$ of monolayer graphene) as a function of angle $\theta$ and exhibits a clear sharp dip at the first magic angle. The fact that the extremely flat bands only occur at certain discrete magic angles indicates that there exists a characteristic condition that breaks the expected monotonic behavior.

It is interesting to examine the developed electron accumulation in the AA region. We present the probability of finding electrons at each atomic site in the lower panels of Fig. \ref{fig:velocity}a-e. The dark and light dots mark the atoms in the first and second layer, respectively, and the dashed dark hexagon marks the Wigner-Seitz cell of the moir\'{e} superlattice. The electron distribution of the zero-energy states at ${K}$ is found to be similar to that integrated over the four red bands in the upper panels. Therefore, we show only the distribution of the four degenerate states at ${K}$, represented by the size of red dots in the lower panels of Fig. \ref{fig:velocity}a-e. As the twist angle is reduced and the AA region clearly develops, the electrons gradually become accumulated in the AA region at the center of the Wigner-Seitz cell. We measure the degree of localization by the integrated fraction of electrons within the dashed blue circle of a radius 0.2$L$, where $L$ is the lattice constant of the moir\'{e} supercell at each angle. This localization measure is found to increase smoothly as the twist angle decreases, as shown by the blue curve in Fig. \ref{fig:velocity}f. As discussed above, this behavior can be explained by the fact that the AA region provides significantly more low-energy states than other stacking patterns. Surprisingly, this smooth behavior is not affected by the presence of the magic angle at all, in contrast to the variation of the Fermi velocity in Fig. \ref{fig:velocity}f. It was proposed previously \cite{PhysRevB.86.155449,tramblydelaissardiere2012} that the series of magic angles are associated with quantization conditions of the confined states in the AA region. However, results in Fig. \ref{fig:velocity}f indicate that the flat band dispersion at the magic angle may not reflect the most localized electronic distribution and that the emergence of the magic angle requires a mechanism other than electron confinement alone.

\subsection*{Connecting AABLG Fermi ring and magic angles in TBG}

As the low-energy electrons are accumulated in the AA region at small angles, apart from the moir\'{e} potential they mostly experience the local lattice potential of AABLG, which has a ring-shaped Fermi surface. This Fermi ring (Fig. \ref{fig:stacking}d) has a particular radius ($Q_F$) in momentum space that is determined by the interlayer interaction. Next we propose a physical explanation for the development of flat bands at the magic angles that is connected to the Fermi ring of AABLG. In particular, we will show that a special situation exists if the TBG Dirac points (in the extended Brillouin zones) fall on the Fermi ring of AABLG. We will then demonstrate that the magic angles occur under these matching conditions and discuss the characteristics of the states associated with the resulting flat bands. 

\begin{table}[h]
\centering
\caption{First magic angle $\theta$ and corresponding moir\'{e} lattice constant $L$ of TBG obtained from tight-binding calculations for different vertical compression levels . These are determined by a commensurate supercell of $m$ that has the smallest Fermi velocity.} \label{tab:TBG_L}
\addtolength{\tabcolsep}{4pt}
\begin{tabular}{ @{}lrrrrrr@{} } 
\toprule
\textbf{compression} & \textbf{0\%} & \textbf{3\%} & \textbf{5\%} & \textbf{10\%} & \textbf{15\%} & \textbf{20\%} \\
\midrule
$m$ & 29 & 26 & 23 & 17 & 13 & 10 \\
$\theta$ ($^\circ$) & 1.12 & 1.25 & 1.41 & 1.89 & 2.45 & 3.15 \\
$L$ (nm) & 12.6 & 11.3 & 10.0 & 7.46 & 5.75 & 4.45 \\
\bottomrule
\end{tabular}
\end{table}

\begin{figure*}[ht]
\centering 
\includegraphics[width=.8\linewidth]{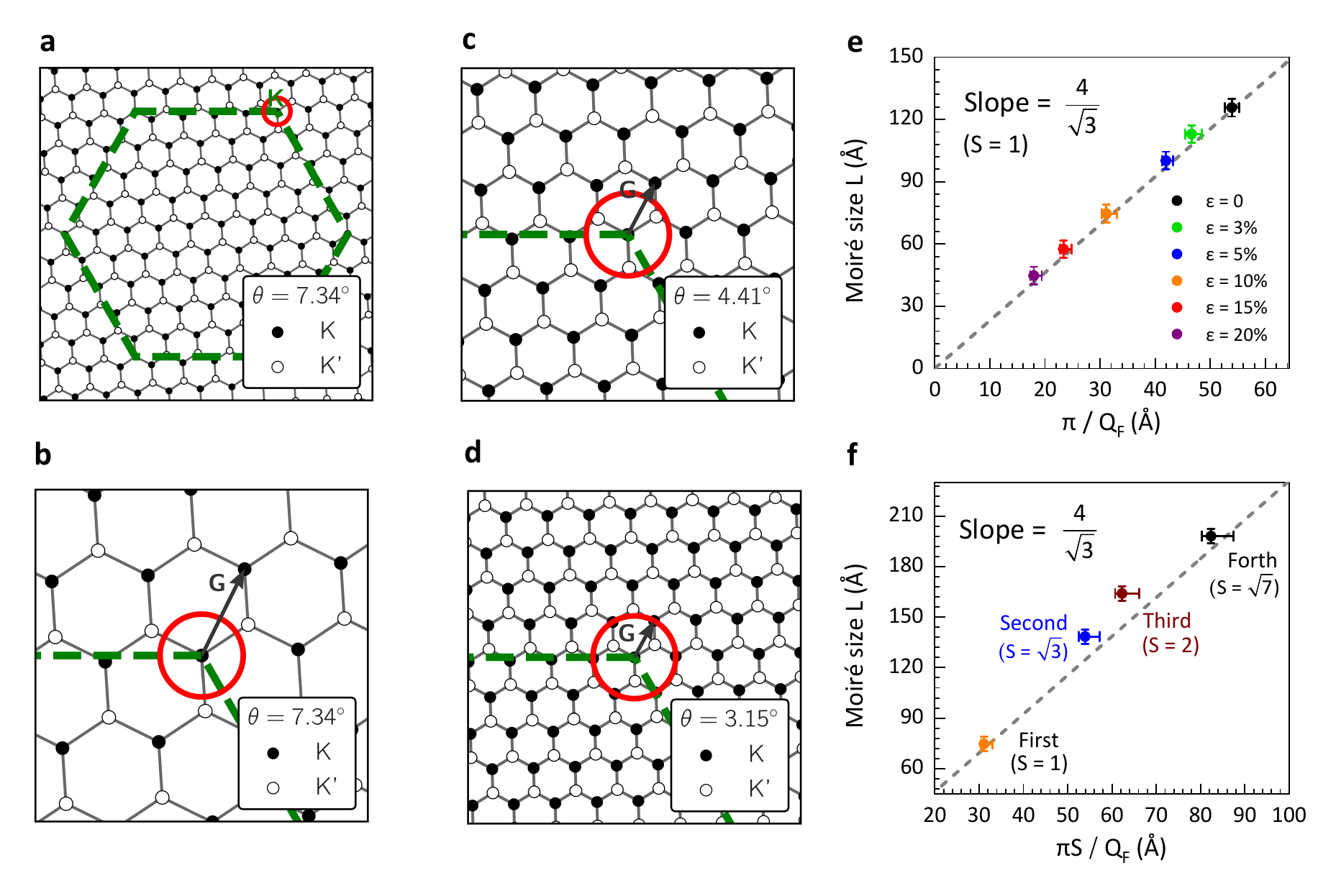}
\caption{$\mid$ {\bf Connection between magic-angle moir\'{e} periodicity and AABLG Fermi ring.} {\bf a-d}, Brillouin zone (dashed green hexagon) and Fermi ring (red circle) of AABLG superimposed on the Brillouin zones (small black hexagons) of TBG at different angles. For a clear illustration, the Fermi ring for AABLG at 20\% compression of the interlayer separation is shown. Both the dashed green hexagon and the red ring are of the same size in {\bf a-d}, but the Brillouin zones of TBG shrink as the moir\'{e} size increases with a decreasing twist angle. The matching condition (see text) is reached at $\theta=3.15^\circ$, which is the magic angle for TBG at 20\% compression. {\bf e}, Correlation between the moir\'{e} size $L$ at the first magic angle ($S = 1$) and the inverse of the radius $Q_F$ of the Fermi ring in AABLG with different interlayer compression. The linear behavior follows the relationship $L=4/\sqrt{3}\times(\pi/Q_F)$ (dashed line). The horizontal error bars on the data points mark the variation in the size of the Fermi ring due to trigonal warping, while the vertical error bars represent the uncertainty in determining the magic angle of TBG since our calculations used a particular collection of commensurate moir\'{e} supercells. {\bf f}, Similar linear connection is found for the second ($S = \sqrt{3}$), third ($S = 2$), and the fourth ($S = \sqrt{7}$) magic angles for the TBG system with 10\% compression.}
\label{fig:match1}
\end{figure*}

The interaction between the layers can be changed by applying uniaxial pressure, namely, by imposing a compressive vertical stress. With different reduced interlayer distances, both the magic angle and the size of the Fermi ring are separately modified. This provides a collection of sample systems that allow us to examine our proposed mechanism systematically by tight-binding calculations. The band dispersion of TBG with different compressed vertical strain values (0\%, 3\%, 5\%, 10\%, 15\%, and 20\%) was evaluated as a function of the twist angle in order to determine their corresponding magic angles. The complete results are listed in Table 1. When the interlayer separation is compressed, the magic angle occurs at a larger angle \cite{PhysRevB.98.085144}. For example, the first magic angle shifts from $\theta$ = 1.12$^\circ$ at 0\% compression to 3.15$^\circ$ at 20\% compression. At the same time, the radius of the Fermi ring in AABLG also enlarges as the bilayer is compressed. For large compression values, a trigonal warping may be found, and an approximated Fermi ring is determined by a circle with the same area. This deviation will be considered in later discussions.

Since the compressed bilayer graphene has a smaller moir\'{e} supercell at the first magic angle, it is easier to graphically visualize this effect in a compressed bilayer system. We plot in Fig. \ref{fig:match1}a-d the 2D Brillouin zone (dashed green hexagon) and Fermi ring (red circle) of AABLG with 20\% vertical compression. Superimposed in the figures are the Brillouin zones (small black hexagons) of TBG at different angles, where the $K$ and $K'$ Dirac points are marked by solid and open circles, respectively. Figure \ref{fig:match1}a shows the configuration for $\theta$ = 7.34$^\circ$, with an enlarged view in Fig. \ref{fig:match1}b. At this angle, the Dirac points of TBG do not seem to be influenced by the AABLG Fermi ring. Note that the Brillouin zone of TBG shrinks as the moir\'{e} size increases with a decreasing twist angle. A special matching condition is found in Fig. \ref{fig:match1}d, where the Fermi ring of AABLG centered at $K$ reaches the first star of Dirac $K$ points in the extended Brillouin zones of TBG. We will show below that this particular matching condition yields the first magic angle in systems we have investigated.

\begin{figure*}[ht]
\centering 
\includegraphics[width=.8\linewidth]{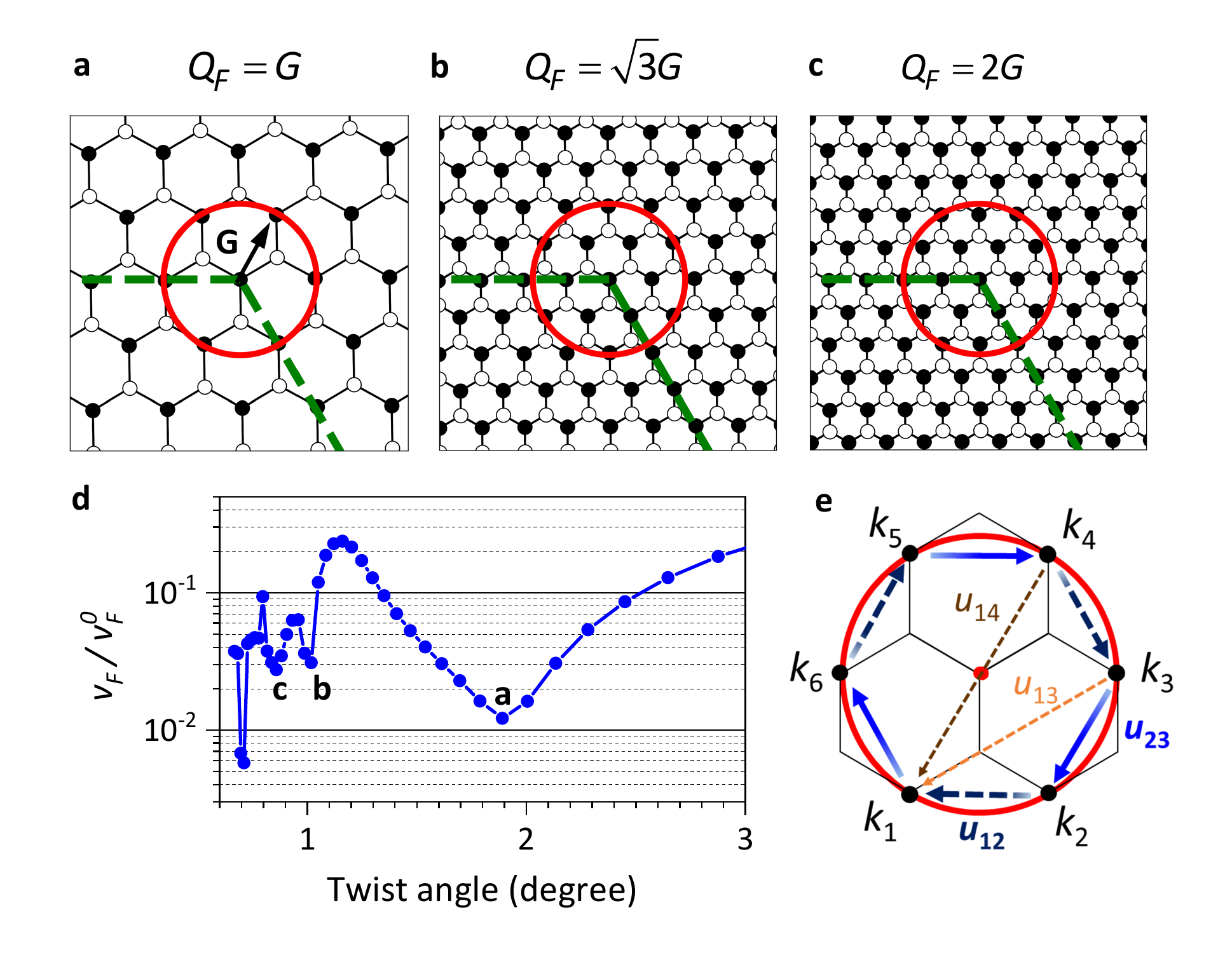}
\caption{$\mid$ {\bf Higher-order magic angles.} {\bf a-c}, Matching conditions between the AABLG Fermi ring (red circle) and the first, second, and third stars of the $K$ points (black filled circles) of TBG in the extended zones (black hexagons). These give rise to the first, second, and third magic angles, respectively. The magnitude of $\bm{G}$, the smallest reciprocal lattice vector of the moir\'{e} supercell decreases in {\bf a-c}. {\bf d}, Calculated (normalized) Fermi velocity as a function of the twist angle for TBG with 10\% vertical compression, showing four dips corresponding to the first four magic angles. {\bf e}, Enlarged matching condition in {\bf a} showing the six $K$ points on the Fermi ring and possible scattering between them via reciprocal lattice vectors of the moir\'{e} supercell.}
\label{fig:match2}
\end{figure*}

We plot in Fig. \ref{fig:match1}e the moir\'{e} supercell size $L$ at the first magic angle found in the calculation (determined by the dip in the calculated Fermi velocity) versus the inverse of $Q_F$ (the radius of the AABLG Fermi ring) for bilayer systems with various compression levels. These two quantities were independently obtained for two different systems by our tight-binding calculations. The horizontal error bars on the data points mark the variation in the size of the Fermi ring due to the trigonal warping: the smallest (largest) radius in the ${K\mathit{\Gamma}}$ (${KM}$) direction gives the upper (lower) bound of $\pi/Q_F$. The vertical error bars represent the uncertainty in determining the exact magic angle, because our calculations used a particular collection of commensurate moir\'{e} supercells. If the largest Fermi velocity dip was found in the calculation for a moir\'{e} supercell with a particular $m$ value, then the moir\'{e} size for the exact magic angle is bounded by those corresponding to $m-1$ and $m+1$ moir\'{e} supercells. As one can see from Fig. \ref{fig:match1}e, an almost perfect linear behavior is found with a slope of $4/\sqrt{3}$. In other words, \textit{the size of the AABLG Fermi ring determines the moir\'{e} periodicity where the magic angle occurs}. And the first magic angle appears when the radius of the Fermi ring matches $G$, the smallest magnitude of nonzero reciprocal lattice vectors of the moir\'{e} supercell, as shown in Fig. \ref{fig:match1}d and Fig. \ref{fig:match2}a. As will be shown below, the critical matching condition is $Q_F = SG$, where $S = 1$ corresponds to the first magic angle.

When the radius of the AABLG Fermi ring matches the second and third stars of the Dirac $K$ points, as shown in Fig. \ref{fig:match2}b and c, we expect to see the second ($S = \sqrt{3}$) and third ($S = 2$) magic angles, respectively. We show in Fig. \ref{fig:match2}d a more complete plot of the calculated Fermi velocity versus $\theta$ for TBG with a 10\% compression, and data up to the fourth magic angle (0.71$^\circ$, $S=\sqrt{7}$) was obtained. Even though the calculation is done for a compression value of 10\%, the physics is the same no matter what compression value (namely, interlayer coupling) is considered. The corresponding moir\'{e} size at the first four magic angles versus $\pi S/Q_F$ is plotted in Fig. \ref{fig:match1}f. Again, a consistent linear behavior is found that matches the relationship

\begin{equation}
    L = \frac{4}{\sqrt{3}} \frac{\pi}{Q_F} S,
\end{equation}
where $S = 1,\sqrt 3 ,2, \sqrt 7$. Given that the minimal length of the reciprocal lattice vector is $G = 4\pi/(\sqrt{3}L)$, $Q_F = S\,G$ is then the criterion in momentum space that determines the occurrence of the series of magic angles. Possible $S$ values are expected to reflect the radius of the stars of $K$ points: $S=1,\sqrt 3 ,2,\sqrt 7 ,3,2\sqrt 3 ,...$. Under these matching conditions, the zero-energy states within AABLG (Fermi ring) and the zero-energy states within TBG (Dirac points) are aligned. 

At small angles, we have $L\approx a/\theta$, where $a$ is the lattice constant of monolayer graphene. Therefore, the magic angles in TBG can be expressed as

\begin{equation}
    \theta_{magic}\approx\frac{Q_F}{\sqrt{3}\,\, l_K}\frac{1}{S}\,,
\end{equation}
where $l_K = 4\pi/(3a)$ is the distance from $\mathit{\Gamma}$ to $K$ for monolayer graphene, and $S$ is not necessarily an integer. 
Our DFT value for $Q_F$ is about 0.055 \AA$^{-1}$. Therefore, the predicted magic angles are 1.07$^\circ$, 0.62$^\circ$, 0.53$^\circ$, 0.40$^\circ$, $\ldots$ for TBG. 
If a tight-binding model considers only the nearest neighbor coupling $t_\bot$ for the interlayer interaction, the AABLG Fermi ring will have a radius of $Q_F = t_\bot /(\hbar v^0_F)$.

\subsection*{Wave functions of the Dirac point at magic angles}

Away from the magic angle, the energy bands in Fig. \ref{fig:velocity} can be explained by the following picture: As the two layers are rotated by an angle in real space, the Dirac cones of the two layers are also rotated with respect to each other in momentum space; the interlayer interaction induces coupling between the two sets of linear bands, giving rise to a reduced slope (Fermi velocity) that is supposed to vary monotonically with the angle. This simple picture is understandable, since each moir\'{e} supercell has regions of different stacking patterns (see Fig. \ref{fig:stacking}a) that cannot influence the energy bands in an abrupt way. On the other hand, when the angle gets small, the relevant low-energy electrons are accumulated in the AA region, so they should be highly influenced by the lattice potential of AABLG. When the matching condition $Q_F = S\,G$ is satisfied at the magic angles, it would be reasonable to construct the wave functions for the Dirac point of TBG from the Bloch states on the AABLG Fermi ring.

We use the first magic angle ($Q_F = G$) as an example. Figure \ref{fig:match2}e shows the detailed geometry with six Dirac points on the AABLG Fermi ring; they could be folded in TBG to the $K$ point at the ring center (marked by a red point). The Bloch states on the AABLG Fermi ring are doubly degenerate and half occupied in a neutral system. Using the nearest-neighbor tight-binding model with an orthogonal basis of $p_z$ orbitals on the $A$ and $B$ sublattices in layer 1 and 2: \{$A_1$, $A_2$, $B_1$, $B_2$\}, these two degenerate states can be chosen as $\left\vert {k}\right\rangle$ $\propto$ $(1\;0\;0\;{-e^{i\phi}})^T$ and $(0\;1\;{-e^{i\phi}}\;0)^T$ with $\phi$ = ${\tan ^{-1}}\left( {{k_y}/{k_x}} \right)$. These two degenerate states will not be coupled with each other by a local moir\'{e} potential, so in the following analysis we make this assumption and deal with one set of them at a time. Each pair of the six $k$ points in Fig. \ref{fig:match2}e are connected by a reciprocal lattice vector of the moir\'{e} supercell, so they will be coupled with each other via the scattering by a 2D moir\'{e} potential $U(\bm{r}) = \Sigma_{\bm{G'}} {U_{\bm{G'}}} e^{i {\bm{G'}}\cdot {\bm{r}}}$, where $\bm{G'}$ is a reciprocal lattice vector of the moir\'{e} superlattice. With the combination of ${C_3}$ (vertical axis) and ${C_2}$ (horizontal axes) rotation symmetries and the fact that the moir\'{e} potential is real ($U_{\bm{G'}}$ = $U^*_{-\bm{G'}}$), the number of independent Fourier coefficients is greatly reduced. We denote that $U_{\bm{G'}=0}$ = $U_0$, $U_{\bm{k_1}-\bm{k_2}}$ = $U_{\bm{k_2}-\bm{k_3}}$ = $\ldots$ = $U_1$, and $U_{\bm{k_1}-\bm{k_3}}$ = $U^*_{\bm{k_2}-\bm{k_4}}$ = $U_{\bm{k_3}-\bm{k_5}}$ = $\ldots$ = $U_2$. $U_1$ is real, and $U_2$ can be made real by choosing the 2D origin at an inversion center.

After including the symmetry of the system and assuming that the long-scale moir\'{e} potential varies little over a single graphene unit cell (see Methods), the perturbation Hamiltonian of the moir\'{e} potential for one set of the wave functions at the six $k$ points ($\bm{k_1}$, $\bm{k_2}$, $\ldots$, $\bm{k_6}$ in Fig. \ref{fig:match2}e) has this form:

\begin{equation}
H = \left( {\begin{array}{*{20}{c}}
{u_{11}}&{{u_{12}}}&{{u_{13}}}&0&{u_{13}^*}&{u_{12}^*}\\
{}&{u_{11}}&{{u_{12}}}&{{e^{i2\pi/3}}{u^*_{13}}}&0&{e^{-i2\pi/3}}{u_{13}}\\
{}&{}&{u_{11}}&{{u_{12}}}&{{u_{13}}}&0\\
{}&{}&{}&{u_{11}}&{{u_{12}}}&{{e^{i2\pi/3}}{u^*_{13}}}\\
{}&{\mathrm{c.}\,\mathrm{c.}}&{}&{}&{u_{11}}&{{u_{12}}}\\
{}&{}&{}&{}&{}&{u_{11}}
\end{array}} \right),
\end{equation}
where ${u_{ij}}$ = $\left\langle {k_i} \right\vert U\left\vert {k_j} \right\rangle$ with the following values:
${u_{11}}$ = $U_0$,
${u_{12}}$ = $(\sqrt{3}/2)\, {U_1}\,{e^{i\pi /6}}$,
${u_{13}}$ = $(1/2)\,U_2\,{e^{i\pi /3}}$, and ${u_{14}}=0$.
This perturbation Hamiltonian turns out to have three doubly degenerate eigenstates. After a constant shift of $U_0 - \operatorname{Re}[U_2]$ to align the energy of the middle state with zero, we have the three eigenvalues:

\begin{equation}
{E_0} = 0\,;\;\;{E_\pm} = \frac{3}{2}\left[\operatorname{Re}[U_2]\pm \sqrt{U_1^2+\left({\operatorname{Im}[U_2]}/\sqrt{3}\right)^2}\, \right]\,.
\end{equation}
In the space of the six $k$ states on the Fermi ring of AABLG, the two degenerate eigenvectors \(v_1\) and \(v_2\) for $E_0$ are:

\begin{equation}
{v_1} = \left( {\begin{array}{*{20}{c}}
0\\
1\\
0\\
{{e^{i2\pi /3}}}\\
0\\
{{e^{i4\pi /3}}}
\end{array}} \right)\,{\mathrm{and}}\;\;{v_2} = \left( {\begin{array}{*{20}{c}}
1\\
0\\
{{e^{i2\pi /3}}}\\
0\\
{{e^{i4\pi /3}}}\\
0
\end{array}} \right)\;,
\end{equation}
where $v_1$ is a linear combination of $k_1$, $k_3$, and $k_5$ states, while $v_2$ is made of $k_2$, $k_4$, and $k_6$ states. Both $v_1$ and $v_2$ are invariant under the $C_3$ rotation and with the same eigenvalue $\omega$. (The value of $\omega$ could be $e^{-i2\pi/3}$ or 1 with the rotation axis chosen at a C atom or the center of a six-atom ring.) The results for the other set of degenerate Bloch states at the six $k$ points are the same. Therefore, we obtain four zero-energy degenerate states at the $K$ point of TBG that all transform in the same way under the $C_3$ rotation. For higher magic angles, the matrix elements in Eq. (3) will involve higher Fourier coefficients of the moir\'{e} potential. Nevertheless, the features of the final eigenvectors will be similar to those for $S$ = 1.

Next we show that the bands turn out to be flat going away from the $K$ point by the $k\cdot p$ perturbation theory. The main $O(k)$ correction in energy away from the $K$ point comes from the $k\cdot p$ intraband coupling among the fourfold degenerate states at $E$=0. These intraband couplings, however, turn out to vanish because all the four eigenstates ($v_l$) at $E$=0 have the same $C_3$ rotation symmetry with the same eigenvalue of $\omega$. This can be seen as follows.
First, we rewrite $k\cdot\hat{p}={{k}_{+}}{{{\hat{p}}}_{-}}+{{k}_{-}}{{{\hat{p}}}_{+}}, $ where ${{k}_{\pm}}=({{k}_{x}}\pm i{{k}_{y}})/\sqrt{2}$ and ${{{\hat{p}}}_{\pm}}=({{{\hat{p}}}_{x}}\pm i{{{\hat{p}}}_{y}})/\sqrt{2}$.
The operator ${\hat{p}}_{\pm}$ transforms as $C_n {\hat{p}}_{\pm} C_n^{\dag}$ = $e^{\mp i2\pi/n} {\hat{p}}_{\pm}$ under the $C_n$ rotation.
Then $\left\langle {{v}_{l}} \right\vert{{\hat{p}}_{+}}\left\vert {{v}_{m}} \right\rangle$ = $\left\langle {{v}_{l}} \right\vert C_{3}^{\dag}{{C}_{3}}\,{{\hat{p}}_{+}}C_{3}^{\dag}{{C}_{3}}\left\vert {{v}_{m}} \right\rangle$ = $\left\langle {{v}_{l}} \right\vert{\omega^*}(e^{-i2\pi/3}{{\hat{p}}_{+}})\omega\left\vert {{v}_{m}} \right\rangle$ = $e^{-i2\pi/3}\left\langle  {{v}_{l}} \right\vert{{{\hat{p}}}_{+}}\left\vert {{v}_{m}} \right\rangle$.
Therefore, $\left\langle  {{v}_{l}} \right\vert{{\hat{p}}_{+}}\left\vert {{v}_{m}} \right\rangle$ = 0, and similarly $\left\langle  {{v}_{l}} \right\vert{{\hat{p}}_{-}}\left\vert {{v}_{m}} \right\rangle$ = 0.
As such, the resonant scattering by the moiré potential regulates the $C_3$ rotation symmetry of the degenerate states at $E$=0, kills the intraband $k\cdot p$ coupling, and thus significantly flattens the bands away from $K$. 

In addition, the net $O({k^2})$ correction from the interband coupling also tends to be small. If we neglect the long-range interaction and set $U_2$=0, we have $E_+$ = ${-E_-}$ and $\sigma H\sigma = {-H}$ with $\sigma$ =  Diag[1,-1,1,-1,1,-1].
Consequently, contributions from the interband coupling cancel each other: the ${E_+}$ bands tend to bend downward the ${E_0}$ band, while the ${E_-}$ bands tend to bend it upward, rendering a net zero correction in energy to the order of $O({k^2})$.

\section{Discussion}

In this work we have identified the following key factors for the occurrence of magic angles in TBG, which may suggest the directions for searching for this interesting phenomenon in other 2D systems.

First, there exists a particular stacking pattern within the the moir\'{e} supercell that attracts low-energy electrons to accumulate. This charge concentration becomes significant at small twist angles, because the region associated with the particular stacking pattern becomes substantial over many bond lengths. This happens in the AA region of TBG that has more low-energy states available compared with regions with other stacking patterns. If this condition does not occur and the electrons spread out throughout the moir\'{e} supercell, the averaging effect will only decrease the bandwidth monotonically as the supercell size increases.

Second, this particular region has its characteristic band structure near the Fermi level that will respond to the perturbing moir\'{e} potential in a particular way only at certain twist angles, where the resulting electronic states can have special features. In the case of TBG, the AA region has a characteristic Fermi ring, and the states on the ring can be coherently coupled when the size of the ring matches the reciprocal lattice vectors of the moir\'{e} supercell ($Q_F=S G$). Therefore, the size of this Fermi ring determines the moir\'{e} periodicity where the series of magic angles occur, and the corresponding TBG wave functions at the Dirac $K$ point are special linear combinations of wave functions on the AABLG Fermi ring.

Third, the characteristic wave functions at the magic angles have certain symmetry properties that are able to induce a flat band dispersion away from the reference $k$ point. This is possible because with a proper symmetry transformation property for $\hat{p}_{\pm}$, the first-order correction within the $k\cdot p$ perturbation theory can vanish. Possible symmetries include inversion, $C_n$ rotations, and reflections. At the magic angles in TBG, the fourfold degenerate $E$=0 states at $K$ have the same eigenvalue under the $C_3$ rotation. The transformation property $C_3 {\hat{p}}_{\pm} C_3^{\dag}$ = $e^{\mp i2\pi/3} {\hat{p}}_{\pm}$ sets the intraband $k\cdot p$ coupling to zero, giving rise to flat bands out of $K$ to the linear order.

The calculations presented in this work did not include the lattice relaxation effect in TBG. Previous theoretical studies have shown that the relaxation pattern in the AA region can simply be described as adding an extra small local rotation around the center of the AA region \cite{PhysRevB.96.075311,PhysRevB.98.224102,PhysRevB.107.075408}. For example, for the case of $\theta= 1.05^{\circ}$, the local rotation angle between the two layers is increased from $1.05^{\circ}$ to $1.63^{\circ}$ upon relaxation \cite{PhysRevB.96.075311}, and for very small $\theta$ values, the relaxed local twist angle at the AA stacking converges to $1.7^{\circ}$ \cite{PhysRevB.98.224102}. As a consequence, the AA stacking remains intact after relaxation, albeit with a reduced extent. Nevertheless, the remaining size is still of the order of a few nanometers for small twist angles (see Fig. 5 in \cite{uchida2014} for $\theta= 0.99^{\circ}$ and Fig. 7 in \cite{PhysRevB.96.075311} for $\theta= 1.05^{\circ}$). This is consistent with the STM topographic images at around  $\theta= 1.1^{\circ}$ that also showed bright spots of AA stacking regions with a size of a few nanometers \cite{Kerelsky2019,Tilak2021}.  Since there still exist sizable AA regions after one takes into account the lattice relaxation, our theory based on the Fermi ring of the AA stacking should remain valid.

A recent paper by Yu \textsl{et al.} \cite{PhysRevB.108.045138} analyzed the low-energy states around the first magic angle and found a band inversion when the twist angle changes across the magic angle. This band inversion at $\Gamma$ was explained by the hybridization between the conduction band and the valance band from different monolayers, and flat bands occur when the band inversion takes place at the magic angle. However, this paper did not address what causes the hybridization (and the consequent band inversion) and, most importantly, why it happens at this particular angle.
Another previous study focusing on an idealized continuum model by switching off the interlayer coupling parameter $w_{AA}$ for the AA stacking yielded perfectly flat bands at multiple magic angles \cite{PhysRevLett.122.106405}. Intriguingly, the wave functions of these flat bands at the magic angles in this particular model were found to be reminiscent of quantum Hall wave functions on the torus. Because the condition $w_{AA}$ = 0 gives rise to perfectly flat bands, it was further speculated that the fundamental features of TBG are mainly connected with the interlayer coupling $w_{AB}$ in the AB stacking. Since this idealized model does not correspond to any realistic graphene systems, it was hoped that one can extend the results to cases with nonzero $w_{AA}$ by treating $w_{AA}$/$w_{AB}$ as a perturbation. However, this may be difficult because $w_{AA}$ and $w_{AB}$ should be of the same order of magnitude in realistic TBG systems. In contrast, the current work focuses on the important effect of the $w_{AA}$ interaction based on the fact that low-energy electrons in TBG are concentrated in the AA region. Our proposed theory is therefore able to fully explain the occurrence and positions of magic angles in terms of the local electronic properties of the AA region.

Recently, Li \textsl{et al.} \cite{PhysRevLett.128.026404} reported a study on a special graphene system consisting of an AA-stacked bilayer on a monolayer with a rotation between them. The so-called ``Dirac magic angles” in that trilayer graphene system have a different definition and correspond to twist angles for which multiple Dirac cones coincide at the same high symmetry points in a moir\'{e} Brillouin zone. However, no flat bands as in TBG were found at any angle, including those Dirac magic angles. Interestingly, under a hypothetical condition that the interaction between the AA-bilayer and the monolayer is completely turned off (in both the AA and AB regions in the moir\'{e} supercell), the origin of Dirac magic angles can simply be traced to a geometric structure of the moir\'{e} lattice in the reciprocal space, as shown in Fig. 2(a) in the paper. A common feature with our work is that the reciprocal lattice structure is used to explain the origin of the two different types of ``magic angles”. However, the relevant ``stars” in the reciprocal space are different. For the AA-bilayer/monolayer, the series is 1, 2, $\sqrt{7}$, $\sqrt{13}$, $\ldots$, and for TBG in our work, the series is 1, $\sqrt{3}$, 2, $\sqrt{7}$, $\ldots$. This is not surprising, since the physics subject is quite different: coexisting Dirac cones in trilayer graphene versus flat bands in TBG. Nevertheless, the similar consideration suggests that if the system contains a sizable region of AA-stacked bilayer graphene, the existence of its Fermi ring in the reciprocal space may create certain matching conditions leading to interesting physics.

In conclusion, this work resolves the long-standing question why magic angles happen in twisted bilayer graphene. The origin of the magic angles turns out to be the ``magic" Fermi ring in AA-stacked bilayer graphene. We have theoretically derived the series of magic angle values.
In addition, we have identified the composition of the one-particle wave functions associated with the flat bands for further many-body investigations.
\section*{\label{sec:method}Methods}
\subsection*{Tight-binding calculations}
To calculate the electronic properties of TBG, we used an extended tight-binding model proposed by S. Fang and E. Kaxiras \cite{PhysRevB.93.235153} with parameters determined from calculations with density functional theory (DFT). This model includes intralayer couplings up to eighth nearest neighbors, and a strong angular dependence of interlayer couplings between atoms in two different layers. The spin-orbit interaction is neglected. The interlayer couplings under uniaxial pressure were also obtained using similar procedures \cite{PhysRevB.98.085144}. The Hamiltonian was then diagonalized by the FEAST eigenvalue solver \cite{PhysRevB.79.115112}, and the eigenvalues and eigenvectors were verified by Mathematica in some cases. The density of states were calculated by the tetrahedral method \cite{pssb.2220540211}, and the slope at the $K$ point (Fermi velocity) was evaluated based on the Hellmann-Feynman theorem \cite{alon2003} as follows. The velocity operator is given by $\vec v(\vec k) = {\partial _{\vec k}}H(\vec k)$, where $\vec k$ is the crystal momentum, and $H(\vec k)$ is the tight-binding Hamiltonian of the moir\'e structure. Due to the fourfold degeneracy at the Dirac point, we formed a velocity matrix ${v_{ij}} = \left\langle i \mid{v}(K)\mid j \right\rangle $, where $i$ and $j$ are labels for the degenerate states at the Dirac point. The eigenvalues of the velocity matrix give the velocities; the one with the smallest absolute value was taken as the velocity result. Since the velocity is almost isotropic, we show the result for the ${K \rightarrow M}$ direction in most cases. When there was uncertainty, the average between ${K\rightarrow M}$ and ${K \rightarrow \Gamma}$ was taken.

\subsection*{Density functional calculations}
We also performed DFT calculations for bilayer graphene with different stacking patterns using the Vienna \textit{Ab initio} Simulation Package (VASP) \cite{kresse1996,kresse1996a}. We used the projector augmented wave (PAW) method \cite{blochl1994,kresse1999} to treat core electrons and the optB86b functional to include the van der Waals correction \cite{klimes2011,roman-perez2009}, with a plane-wave cutoff energy of 400 eV. The Monkhorst-Pack k-point mesh of 21$\times$21$\times$1 was used in the self-consistent calculations for the 1$\times$1 unit cell. The calculation of the density of states used a more denser 91$\times$91$\times$1 grid.

\subsection*{Matrix elements of the moir\'e potential}
The Bloch-state components constructed from the local orbitals at any of the four sites $\left\{ {{A_1},{A_2},{B_1},{B_2}} \right\}$ in the unit cell of AABLG have the form:

\begin{equation*}
\left\langle \bm{r} \right\vert\left. {{\alpha _i}} \right\rangle  = \frac{1}{{\sqrt {{N_{\alpha ,i}}} }}\sum\limits_{\bm{R_{\alpha ,j}}} {e^{i\bm{k} \cdot \bm{R_{\alpha,j}}}} {\zeta _{{p_z}}}(\bm{r} - \bm{R_{\alpha ,j}} - {\delta _{i,2}}c\hat z)\;,
\end{equation*}
where $\bm{R_{\alpha,j}}$ is the in-plane position vector for sublattice $\alpha$ in the $j{\rm{th}}$ unit cell; $c$ is the interlayer separation, $i$ refers to the layer index \{1,2\}, and ${\delta _{i,2}}$ = 1 if $i$=2. \({\zeta _{{p_z}}}(\bm{r})\) describes the orthonormal ${p_z}$ atomic orbital, and ${N_{\alpha ,i}}$ = $N$ is the total number of atoms pertaining to sublattice $\alpha$ and layer $i$ in AABLG. For the six $\bm{k}$ vectors on the Fermi ring in Fig. \ref{fig:match2}e, there are two degenerate states, and the wave functions can be chosen as $\left\vert {k}\right\rangle$ $\propto$ $(1\;0\;0\;{-e^{i\phi}})^T$ and $(0\;1\;{-e^{i\phi}}\;0)^T$ with $\phi$ = ${\tan ^{-1}}\left( {{k_y}/{k_x}} \right)$. These two states will not be coupled with each other by the local moir\'{e} potential, so we only have to deal with one of them at a time. Both of them are eigenstates of the $C_3$ rotation.

The ${C_3}$ rotation symmetry exists for both TBG [and thus the moir\'e potential $U(\bm{r})$] and AABLG. It connects the matrix elements ${u_{ij}}$ = $\left\langle {k_i} \right\vert U\left\vert {k_j} \right\rangle$, for example, $\left\langle {{k_1}} \right\vert U\left\vert {{k_2}} \right\rangle$ = $\left\langle {{k_1}} \right\vert C_3^{\dag} (C_3 U C_3^{\dag}) {C_3}\left\vert {{k_2}} \right\rangle$ = $\left\langle {{k_3}} \right\vert U \left\vert {{k_4}} \right\rangle$. Therefore, we have (as shown in Fig. \ref{fig:match2}e) ${u_{12}} = {u_{34}} = {u_{56}}$ (navy dashed lines); ${u_{23}} = {u_{45}} = {u_{61}}$ (blue lines); ${u_{13}} = {u_{35}} = {u_{51}}$ (orange dashed line); ${u_{24}} = {u_{46}} = {u_{62}}$; and ${u_{14}} = {u_{36}} = {u_{52}}$ (brown dashed line). As a consequence, the $C_3$ symmetry reduces the number of variables from $\left( {{6^2} - 6} \right)/2 = 15$ to 5, which are ${u_{12}},{u_{23}},{u_{13}},{u_{24}},{\rm{ and }}{u_{14}}$.

Since the moir\'e supercell is much larger than the unit cell of graphene, we can assume that the effective moir\'e potential varies little within one unit cell of AABLG. Therefore, we have the following approximation:
\begin{align*}
\int {\mathrm{d^3}{r}} \zeta_{{p_z}}^*(\bm{r} - \bm{R_{\alpha,j}} - {\delta _{i,2}}c\hat z)\,U(\bm{r})\,&{\zeta_{{p_z}}}(\bm{r} - \bm{R_{\alpha ',j'}} - {\delta _{i',2}}c\hat z)\\
&\approx U(\bm{R_{\alpha ,j}}){\delta _{i,i'}}{\delta _{j,j'}}{\delta _{\alpha ,\alpha '}}\,.   
\end{align*}
Therefore, the matrix elements of the 2D moir\'e potential ${u_{lm}} = \left\langle {{k_l}} \right\vert U\left\vert {{k_m}} \right\rangle$
for six $\left\vert {k}\right\rangle$ $\propto$ $(1\;0\;0\;{-e^{i\phi}})^T$ are:
\begin{equation*}{}
\begin{split}
{u_{lm}} & \approx\frac{1}{{2N}}\left[ \sum\limits_{{R_{A,j}}} {{e^{i({\bm{k_m} - \bm{k_l}}) \cdot \bm{R_{A,j}}}}} \,U(\bm{R_{A,j}}) \right. \\
& \left. \hspace{1.4cm} + {e^{i({\phi _m}-{\phi_l})}} \sum\limits_{{R_{B,j}}} {{e^{i(\bm{k_m} - \bm{k_l}) \cdot \bm{R_{B,j}}}}} \,U(\bm{R_{B,j}}) \right]\,.
\end{split}
\end{equation*}
This can be expressed as:
\begin{equation*}{}
\begin{split}
{u_{lm}} & \approx\frac{1}{{2A}}\left( {1 + {e^{i({\phi _m} - {\phi _l})}}} \right)\int {e^{i(\bm{k_m} - \bm{k_l}) \cdot \bm{r}}U(\bm{r})\mathrm{d^2}{r}} \\
& = \frac{1}{2}\left( {1 + {e^{i({\phi _m} - {\phi _l})}}} \right){U_{\bm{k_l} - \bm{k_m}}}\,,
\end{split}
\end{equation*}
where $A$ denotes the whole area of the TBG. 
With the combination of ${C_3}$ and ${C_2}$ rotation symmetries and the fact that the moir\'{e} potential is real ($U_{\bm{G'}}$ = $U^*_{-\bm{G'}}$), we have $U_{\bm{k_1}-\bm{k_2}}$ = $U_{\bm{k_2}-\bm{k_3}}$ = $\ldots$ = $U_1$, and $U_{\bm{k_1}-\bm{k_3}}$ = $U^*_{\bm{k_2}-\bm{k_4}}$ = $U_{\bm{k_3}-\bm{k_5}}$ = $\ldots$ = $U_2$. $U_1$ is real, and $U_2$ is generally complex but can be made real by choosing the 2D origin at an inversion center. Given that $\phi_2$ - $\phi_1$ = $\pi/3$, $\phi_3$ - $\phi_1$ = $2\pi/3$, $\phi_4$ - $\phi_1$ = $\pi$, $(1+e^{i\theta})/2$ = $\cos(\theta/2)e^{i\theta/2}$, and $u_{lm}$ = $u_{ml}^*$, we then have
${u_{11}}$ = $U_{G=0}$,
${u_{12}}$ = $(\sqrt{3}/2)\, U_1\,{e^{i\pi /6}}$ = ${u_{23}}$,
${u_{13}}$ = $(1/2)\,U_2\,{e^{i\pi /3}}$, ${u_{24}}$ = $e^{i2\pi/3}u_{13}^*$ and ${u_{14}}=0$.

\begin{acknowledgments}
We gratefully acknowledge Li-Feng Yen and Po-Tung Fang for sharing the tight-binding codes and for helping with technical matters. Helpful discussions with Wen-Ying Ruan, Chi-Ruei Pan, Jie-Cheng Chen, Wei-En Tseng, and Martin Callsen are acknowledged. This work is supported by Academia Sinica, Taiwan.
\end{acknowledgments}


\bibliography{bib_arXiv}

\begin{thebibliography}{61}%
\makeatletter
\providecommand \@ifxundefined [1]{%
 \@ifx{#1\undefined}
}%
\providecommand \@ifnum [1]{%
 \ifnum #1\expandafter \@firstoftwo
 \else \expandafter \@secondoftwo
 \fi
}%
\providecommand \@ifx [1]{%
 \ifx #1\expandafter \@firstoftwo
 \else \expandafter \@secondoftwo
 \fi
}%
\providecommand \natexlab [1]{#1}%
\providecommand \enquote  [1]{``#1''}%
\providecommand \bibnamefont  [1]{#1}%
\providecommand \bibfnamefont [1]{#1}%
\providecommand \citenamefont [1]{#1}%
\providecommand \href@noop [0]{\@secondoftwo}%
\providecommand \href [0]{\begingroup \@sanitize@url \@href}%
\providecommand \@href[1]{\@@startlink{#1}\@@href}%
\providecommand \@@href[1]{\endgroup#1\@@endlink}%
\providecommand \@sanitize@url [0]{\catcode `\\12\catcode `\$12\catcode `\&12\catcode `\#12\catcode `\^12\catcode `\_12\catcode `\%12\relax}%
\providecommand \@@startlink[1]{}%
\providecommand \@@endlink[0]{}%
\providecommand \url  [0]{\begingroup\@sanitize@url \@url }%
\providecommand \@url [1]{\endgroup\@href {#1}{\urlprefix }}%
\providecommand \urlprefix  [0]{URL }%
\providecommand \Eprint [0]{\href }%
\providecommand \doibase [0]{https://doi.org/}%
\providecommand \selectlanguage [0]{\@gobble}%
\providecommand \bibinfo  [0]{\@secondoftwo}%
\providecommand \bibfield  [0]{\@secondoftwo}%
\providecommand \translation [1]{[#1]}%
\providecommand \BibitemOpen [0]{}%
\providecommand \bibitemStop [0]{}%
\providecommand \bibitemNoStop [0]{.\EOS\space}%
\providecommand \EOS [0]{\spacefactor3000\relax}%
\providecommand \BibitemShut  [1]{\csname bibitem#1\endcsname}%
\let\auto@bib@innerbib\@empty
\bibitem [{\citenamefont {Shallcross}\ \emph {et~al.}(2010)\citenamefont {Shallcross}, \citenamefont {Sharma}, \citenamefont {Kandelaki},\ and\ \citenamefont {Pankratov}}]{PhysRevB.81.165105}%
  \BibitemOpen
  \bibfield  {author} {\bibinfo {author} {\bibfnamefont {S.}~\bibnamefont {Shallcross}}, \bibinfo {author} {\bibfnamefont {S.}~\bibnamefont {Sharma}}, \bibinfo {author} {\bibfnamefont {E.}~\bibnamefont {Kandelaki}},\ and\ \bibinfo {author} {\bibfnamefont {O.~A.}\ \bibnamefont {Pankratov}},\ }\bibfield  {title} {\bibinfo {title} {Electronic structure of turbostratic graphene},\ }\href {https://doi.org/10.1103/PhysRevB.81.165105} {\bibfield  {journal} {\bibinfo  {journal} {Phys. Rev. B}\ }\textbf {\bibinfo {volume} {81}},\ \bibinfo {pages} {165105} (\bibinfo {year} {2010})}\BibitemShut {NoStop}%
\bibitem [{\citenamefont {Lopes~dos Santos}\ \emph {et~al.}(2012)\citenamefont {Lopes~dos Santos}, \citenamefont {Peres},\ and\ \citenamefont {Castro~Neto}}]{PhysRevB.86.155449}%
  \BibitemOpen
  \bibfield  {author} {\bibinfo {author} {\bibfnamefont {J.~M.~B.}\ \bibnamefont {Lopes~dos Santos}}, \bibinfo {author} {\bibfnamefont {N.~M.~R.}\ \bibnamefont {Peres}},\ and\ \bibinfo {author} {\bibfnamefont {A.~H.}\ \bibnamefont {Castro~Neto}},\ }\bibfield  {title} {\bibinfo {title} {Continuum model of the twisted graphene bilayer},\ }\href {https://doi.org/10.1103/PhysRevB.86.155449} {\bibfield  {journal} {\bibinfo  {journal} {Phys. Rev. B}\ }\textbf {\bibinfo {volume} {86}},\ \bibinfo {pages} {155449} (\bibinfo {year} {2012})}\BibitemShut {NoStop}%
\bibitem [{\citenamefont {Carr}\ \emph {et~al.}(2017)\citenamefont {Carr}, \citenamefont {Massatt}, \citenamefont {Fang}, \citenamefont {Cazeaux}, \citenamefont {Luskin},\ and\ \citenamefont {Kaxiras}}]{PhysRevB.95.075420}%
  \BibitemOpen
  \bibfield  {author} {\bibinfo {author} {\bibfnamefont {S.}~\bibnamefont {Carr}}, \bibinfo {author} {\bibfnamefont {D.}~\bibnamefont {Massatt}}, \bibinfo {author} {\bibfnamefont {S.}~\bibnamefont {Fang}}, \bibinfo {author} {\bibfnamefont {P.}~\bibnamefont {Cazeaux}}, \bibinfo {author} {\bibfnamefont {M.}~\bibnamefont {Luskin}},\ and\ \bibinfo {author} {\bibfnamefont {E.}~\bibnamefont {Kaxiras}},\ }\bibfield  {title} {\bibinfo {title} {Twistronics: Manipulating the electronic properties of two-dimensional layered structures through their twist angle},\ }\href {https://doi.org/10.1103/PhysRevB.95.075420} {\bibfield  {journal} {\bibinfo  {journal} {Phys. Rev. B}\ }\textbf {\bibinfo {volume} {95}},\ \bibinfo {pages} {075420} (\bibinfo {year} {2017})}\BibitemShut {NoStop}%
\bibitem [{\citenamefont {Lopes~dos Santos}\ \emph {et~al.}(2007)\citenamefont {Lopes~dos Santos}, \citenamefont {Peres},\ and\ \citenamefont {Castro~Neto}}]{PhysRevLett.99.256802}%
  \BibitemOpen
  \bibfield  {author} {\bibinfo {author} {\bibfnamefont {J.~M.~B.}\ \bibnamefont {Lopes~dos Santos}}, \bibinfo {author} {\bibfnamefont {N.~M.~R.}\ \bibnamefont {Peres}},\ and\ \bibinfo {author} {\bibfnamefont {A.~H.}\ \bibnamefont {Castro~Neto}},\ }\bibfield  {title} {\bibinfo {title} {Graphene bilayer with a twist: Electronic structure},\ }\href {https://doi.org/10.1103/PhysRevLett.99.256802} {\bibfield  {journal} {\bibinfo  {journal} {Phys. Rev. Lett.}\ }\textbf {\bibinfo {volume} {99}},\ \bibinfo {pages} {256802} (\bibinfo {year} {2007})}\BibitemShut {NoStop}%
\bibitem [{\citenamefont {Bistritzer}\ and\ \citenamefont {MacDonald}(2011)}]{Bistritzer12233}%
  \BibitemOpen
  \bibfield  {author} {\bibinfo {author} {\bibfnamefont {R.}~\bibnamefont {Bistritzer}}\ and\ \bibinfo {author} {\bibfnamefont {A.~H.}\ \bibnamefont {MacDonald}},\ }\bibfield  {title} {\bibinfo {title} {Moir{\'e} bands in twisted double-layer graphene},\ }\href {https://doi.org/10.1073/pnas.1108174108} {\bibfield  {journal} {\bibinfo  {journal} {Proceedings of the National Academy of Sciences}\ }\textbf {\bibinfo {volume} {108}},\ \bibinfo {pages} {12233} (\bibinfo {year} {2011})}\BibitemShut {NoStop}%
\bibitem [{\citenamefont {Su\'arez~Morell}\ \emph {et~al.}(2010)\citenamefont {Su\'arez~Morell}, \citenamefont {Correa}, \citenamefont {Vargas}, \citenamefont {Pacheco},\ and\ \citenamefont {Barticevic}}]{PhysRevB.82.121407}%
  \BibitemOpen
  \bibfield  {author} {\bibinfo {author} {\bibfnamefont {E.}~\bibnamefont {Su\'arez~Morell}}, \bibinfo {author} {\bibfnamefont {J.~D.}\ \bibnamefont {Correa}}, \bibinfo {author} {\bibfnamefont {P.}~\bibnamefont {Vargas}}, \bibinfo {author} {\bibfnamefont {M.}~\bibnamefont {Pacheco}},\ and\ \bibinfo {author} {\bibfnamefont {Z.}~\bibnamefont {Barticevic}},\ }\bibfield  {title} {\bibinfo {title} {Flat bands in slightly twisted bilayer graphene: Tight-binding calculations},\ }\href {https://doi.org/10.1103/PhysRevB.82.121407} {\bibfield  {journal} {\bibinfo  {journal} {Phys. Rev. B}\ }\textbf {\bibinfo {volume} {82}},\ \bibinfo {pages} {121407} (\bibinfo {year} {2010})}\BibitemShut {NoStop}%
\bibitem [{\citenamefont {{Trambly de Laissardi{\`e}re}}\ \emph {et~al.}(2010)\citenamefont {{Trambly de Laissardi{\`e}re}}, \citenamefont {Mayou},\ and\ \citenamefont {Magaud}}]{tramblydelaissardiere2010}%
  \BibitemOpen
  \bibfield  {author} {\bibinfo {author} {\bibfnamefont {G.}~\bibnamefont {{Trambly de Laissardi{\`e}re}}}, \bibinfo {author} {\bibfnamefont {D.}~\bibnamefont {Mayou}},\ and\ \bibinfo {author} {\bibfnamefont {L.}~\bibnamefont {Magaud}},\ }\bibfield  {title} {\bibinfo {title} {Localization of {{Dirac Electrons}} in {{Rotated Graphene Bilayers}}},\ }\href {https://doi.org/10.1021/nl902948m} {\bibfield  {journal} {\bibinfo  {journal} {Nano Lett.}\ }\textbf {\bibinfo {volume} {10}},\ \bibinfo {pages} {804} (\bibinfo {year} {2010})}\BibitemShut {NoStop}%
\bibitem [{\citenamefont {{Trambly de Laissardi{\`e}re}}\ \emph {et~al.}(2012)\citenamefont {{Trambly de Laissardi{\`e}re}}, \citenamefont {Mayou},\ and\ \citenamefont {Magaud}}]{tramblydelaissardiere2012}%
  \BibitemOpen
  \bibfield  {author} {\bibinfo {author} {\bibfnamefont {G.}~\bibnamefont {{Trambly de Laissardi{\`e}re}}}, \bibinfo {author} {\bibfnamefont {D.}~\bibnamefont {Mayou}},\ and\ \bibinfo {author} {\bibfnamefont {L.}~\bibnamefont {Magaud}},\ }\bibfield  {title} {\bibinfo {title} {Numerical studies of confined states in rotated bilayers of graphene},\ }\href {https://doi.org/10.1103/PhysRevB.86.125413} {\bibfield  {journal} {\bibinfo  {journal} {Phys. Rev. B}\ }\textbf {\bibinfo {volume} {86}},\ \bibinfo {pages} {125413} (\bibinfo {year} {2012})}\BibitemShut {NoStop}%
\bibitem [{\citenamefont {Uchida}\ \emph {et~al.}(2014)\citenamefont {Uchida}, \citenamefont {Furuya}, \citenamefont {Iwata},\ and\ \citenamefont {Oshiyama}}]{uchida2014}%
  \BibitemOpen
  \bibfield  {author} {\bibinfo {author} {\bibfnamefont {K.}~\bibnamefont {Uchida}}, \bibinfo {author} {\bibfnamefont {S.}~\bibnamefont {Furuya}}, \bibinfo {author} {\bibfnamefont {J.-I.}\ \bibnamefont {Iwata}},\ and\ \bibinfo {author} {\bibfnamefont {A.}~\bibnamefont {Oshiyama}},\ }\bibfield  {title} {\bibinfo {title} {Atomic corrugation and electron localization due to moir{\'e} patterns in twisted bilayer graphenes},\ }\href {https://doi.org/10.1103/PhysRevB.90.155451} {\bibfield  {journal} {\bibinfo  {journal} {Phys. Rev. B}\ }\textbf {\bibinfo {volume} {90}},\ \bibinfo {pages} {155451} (\bibinfo {year} {2014})}\BibitemShut {NoStop}%
\bibitem [{\citenamefont {Cao}\ \emph {et~al.}(2016)\citenamefont {Cao}, \citenamefont {Luo}, \citenamefont {Fatemi}, \citenamefont {Fang}, \citenamefont {{Sanchez-Yamagishi}}, \citenamefont {Watanabe}, \citenamefont {Taniguchi}, \citenamefont {Kaxiras},\ and\ \citenamefont {{Jarillo-Herrero}}}]{cao2016}%
  \BibitemOpen
  \bibfield  {author} {\bibinfo {author} {\bibfnamefont {Y.}~\bibnamefont {Cao}}, \bibinfo {author} {\bibfnamefont {J.~Y.}\ \bibnamefont {Luo}}, \bibinfo {author} {\bibfnamefont {V.}~\bibnamefont {Fatemi}}, \bibinfo {author} {\bibfnamefont {S.}~\bibnamefont {Fang}}, \bibinfo {author} {\bibfnamefont {J.~D.}\ \bibnamefont {{Sanchez-Yamagishi}}}, \bibinfo {author} {\bibfnamefont {K.}~\bibnamefont {Watanabe}}, \bibinfo {author} {\bibfnamefont {T.}~\bibnamefont {Taniguchi}}, \bibinfo {author} {\bibfnamefont {E.}~\bibnamefont {Kaxiras}},\ and\ \bibinfo {author} {\bibfnamefont {P.}~\bibnamefont {{Jarillo-Herrero}}},\ }\bibfield  {title} {\bibinfo {title} {Superlattice-{{Induced Insulating States}} and {{Valley-Protected Orbits}} in {{Twisted Bilayer Graphene}}},\ }\href {https://doi.org/10.1103/PhysRevLett.117.116804} {\bibfield  {journal} {\bibinfo  {journal} {Phys. Rev. Lett.}\ }\textbf {\bibinfo {volume} {117}},\ \bibinfo {pages} {116804} (\bibinfo {year} {2016})}\BibitemShut {NoStop}%
\bibitem [{\citenamefont {Kim}\ \emph {et~al.}(2016)\citenamefont {Kim}, \citenamefont {Yankowitz}, \citenamefont {Fallahazad}, \citenamefont {Kang}, \citenamefont {Movva}, \citenamefont {Huang}, \citenamefont {Larentis}, \citenamefont {Corbet}, \citenamefont {Taniguchi}, \citenamefont {Watanabe}, \citenamefont {Banerjee}, \citenamefont {LeRoy},\ and\ \citenamefont {Tutuc}}]{kim2016}%
  \BibitemOpen
  \bibfield  {author} {\bibinfo {author} {\bibfnamefont {K.}~\bibnamefont {Kim}}, \bibinfo {author} {\bibfnamefont {M.}~\bibnamefont {Yankowitz}}, \bibinfo {author} {\bibfnamefont {B.}~\bibnamefont {Fallahazad}}, \bibinfo {author} {\bibfnamefont {S.}~\bibnamefont {Kang}}, \bibinfo {author} {\bibfnamefont {H.~C.~P.}\ \bibnamefont {Movva}}, \bibinfo {author} {\bibfnamefont {S.}~\bibnamefont {Huang}}, \bibinfo {author} {\bibfnamefont {S.}~\bibnamefont {Larentis}}, \bibinfo {author} {\bibfnamefont {C.~M.}\ \bibnamefont {Corbet}}, \bibinfo {author} {\bibfnamefont {T.}~\bibnamefont {Taniguchi}}, \bibinfo {author} {\bibfnamefont {K.}~\bibnamefont {Watanabe}}, \bibinfo {author} {\bibfnamefont {S.~K.}\ \bibnamefont {Banerjee}}, \bibinfo {author} {\bibfnamefont {B.~J.}\ \bibnamefont {LeRoy}},\ and\ \bibinfo {author} {\bibfnamefont {E.}~\bibnamefont {Tutuc}},\ }\bibfield  {title} {\bibinfo {title} {Van der {{Waals Heterostructures}} with {{High Accuracy Rotational Alignment}}},\ }\href
  {https://doi.org/10.1021/acs.nanolett.5b05263} {\bibfield  {journal} {\bibinfo  {journal} {Nano Lett.}\ }\textbf {\bibinfo {volume} {16}},\ \bibinfo {pages} {1989} (\bibinfo {year} {2016})}\BibitemShut {NoStop}%
\bibitem [{\citenamefont {Kim}\ \emph {et~al.}(2017)\citenamefont {Kim}, \citenamefont {DaSilva}, \citenamefont {Huang}, \citenamefont {Fallahazad}, \citenamefont {Larentis}, \citenamefont {Taniguchi}, \citenamefont {Watanabe}, \citenamefont {LeRoy}, \citenamefont {MacDonald},\ and\ \citenamefont {Tutuc}}]{kim2017}%
  \BibitemOpen
  \bibfield  {author} {\bibinfo {author} {\bibfnamefont {K.}~\bibnamefont {Kim}}, \bibinfo {author} {\bibfnamefont {A.}~\bibnamefont {DaSilva}}, \bibinfo {author} {\bibfnamefont {S.}~\bibnamefont {Huang}}, \bibinfo {author} {\bibfnamefont {B.}~\bibnamefont {Fallahazad}}, \bibinfo {author} {\bibfnamefont {S.}~\bibnamefont {Larentis}}, \bibinfo {author} {\bibfnamefont {T.}~\bibnamefont {Taniguchi}}, \bibinfo {author} {\bibfnamefont {K.}~\bibnamefont {Watanabe}}, \bibinfo {author} {\bibfnamefont {B.~J.}\ \bibnamefont {LeRoy}}, \bibinfo {author} {\bibfnamefont {A.~H.}\ \bibnamefont {MacDonald}},\ and\ \bibinfo {author} {\bibfnamefont {E.}~\bibnamefont {Tutuc}},\ }\bibfield  {title} {\bibinfo {title} {Tunable moir\'e bands and strong correlations in small-twist-angle bilayer graphene},\ }\href {https://doi.org/10.1073/pnas.1620140114} {\bibfield  {journal} {\bibinfo  {journal} {PNAS}\ }\textbf {\bibinfo {volume} {114}},\ \bibinfo {pages} {3364} (\bibinfo {year} {2017})}\BibitemShut {NoStop}%
\bibitem [{\citenamefont {Frisenda}\ \emph {et~al.}(2018)\citenamefont {Frisenda}, \citenamefont {{Navarro-Moratalla}}, \citenamefont {Gant}, \citenamefont {Lara}, \citenamefont {{Jarillo-Herrero}}, \citenamefont {Gorbachev},\ and\ \citenamefont {{Castellanos-Gomez}}}]{frisenda2018}%
  \BibitemOpen
  \bibfield  {author} {\bibinfo {author} {\bibfnamefont {R.}~\bibnamefont {Frisenda}}, \bibinfo {author} {\bibfnamefont {E.}~\bibnamefont {{Navarro-Moratalla}}}, \bibinfo {author} {\bibfnamefont {P.}~\bibnamefont {Gant}}, \bibinfo {author} {\bibfnamefont {D.~P.~D.}\ \bibnamefont {Lara}}, \bibinfo {author} {\bibfnamefont {P.}~\bibnamefont {{Jarillo-Herrero}}}, \bibinfo {author} {\bibfnamefont {R.~V.}\ \bibnamefont {Gorbachev}},\ and\ \bibinfo {author} {\bibfnamefont {A.}~\bibnamefont {{Castellanos-Gomez}}},\ }\bibfield  {title} {\bibinfo {title} {Recent progress in the assembly of nanodevices and van der {{Waals}} heterostructures by deterministic placement of {{2D}} materials},\ }\href {https://doi.org/10.1039/C7CS00556C} {\bibfield  {journal} {\bibinfo  {journal} {Chem. Soc. Rev.}\ }\textbf {\bibinfo {volume} {47}},\ \bibinfo {pages} {53} (\bibinfo {year} {2018})}\BibitemShut {NoStop}%
\bibitem [{\citenamefont {{Ribeiro-Palau}}\ \emph {et~al.}(2018)\citenamefont {{Ribeiro-Palau}}, \citenamefont {Zhang}, \citenamefont {Watanabe}, \citenamefont {Taniguchi}, \citenamefont {Hone},\ and\ \citenamefont {Dean}}]{ribeiro-palau2018}%
  \BibitemOpen
  \bibfield  {author} {\bibinfo {author} {\bibfnamefont {R.}~\bibnamefont {{Ribeiro-Palau}}}, \bibinfo {author} {\bibfnamefont {C.}~\bibnamefont {Zhang}}, \bibinfo {author} {\bibfnamefont {K.}~\bibnamefont {Watanabe}}, \bibinfo {author} {\bibfnamefont {T.}~\bibnamefont {Taniguchi}}, \bibinfo {author} {\bibfnamefont {J.}~\bibnamefont {Hone}},\ and\ \bibinfo {author} {\bibfnamefont {C.~R.}\ \bibnamefont {Dean}},\ }\bibfield  {title} {\bibinfo {title} {Twistable electronics with dynamically rotatable heterostructures},\ }\href {https://doi.org/10.1126/science.aat6981} {\bibfield  {journal} {\bibinfo  {journal} {Science}\ }\textbf {\bibinfo {volume} {361}},\ \bibinfo {pages} {690} (\bibinfo {year} {2018})}\BibitemShut {NoStop}%
\bibitem [{\citenamefont {Cao}\ \emph {et~al.}(2018{\natexlab{a}})\citenamefont {Cao}, \citenamefont {Fatemi}, \citenamefont {Demir}, \citenamefont {Fang}, \citenamefont {Tomarken}, \citenamefont {Luo}, \citenamefont {{Sanchez-Yamagishi}}, \citenamefont {Watanabe}, \citenamefont {Taniguchi}, \citenamefont {Kaxiras}, \citenamefont {Ashoori},\ and\ \citenamefont {{Jarillo-Herrero}}}]{cao2018a}%
  \BibitemOpen
  \bibfield  {author} {\bibinfo {author} {\bibfnamefont {Y.}~\bibnamefont {Cao}}, \bibinfo {author} {\bibfnamefont {V.}~\bibnamefont {Fatemi}}, \bibinfo {author} {\bibfnamefont {A.}~\bibnamefont {Demir}}, \bibinfo {author} {\bibfnamefont {S.}~\bibnamefont {Fang}}, \bibinfo {author} {\bibfnamefont {S.~L.}\ \bibnamefont {Tomarken}}, \bibinfo {author} {\bibfnamefont {J.~Y.}\ \bibnamefont {Luo}}, \bibinfo {author} {\bibfnamefont {J.~D.}\ \bibnamefont {{Sanchez-Yamagishi}}}, \bibinfo {author} {\bibfnamefont {K.}~\bibnamefont {Watanabe}}, \bibinfo {author} {\bibfnamefont {T.}~\bibnamefont {Taniguchi}}, \bibinfo {author} {\bibfnamefont {E.}~\bibnamefont {Kaxiras}}, \bibinfo {author} {\bibfnamefont {R.~C.}\ \bibnamefont {Ashoori}},\ and\ \bibinfo {author} {\bibfnamefont {P.}~\bibnamefont {{Jarillo-Herrero}}},\ }\bibfield  {title} {\bibinfo {title} {Correlated insulator behaviour at half-filling in magic-angle graphene superlattices},\ }\href {https://doi.org/10.1038/nature26154} {\bibfield  {journal} {\bibinfo
  {journal} {Nature}\ }\textbf {\bibinfo {volume} {556}},\ \bibinfo {pages} {80} (\bibinfo {year} {2018}{\natexlab{a}})}\BibitemShut {NoStop}%
\bibitem [{\citenamefont {Cao}\ \emph {et~al.}(2018{\natexlab{b}})\citenamefont {Cao}, \citenamefont {Fatemi}, \citenamefont {Fang}, \citenamefont {Watanabe}, \citenamefont {Taniguchi}, \citenamefont {Kaxiras},\ and\ \citenamefont {{Jarillo-Herrero}}}]{cao2018b}%
  \BibitemOpen
  \bibfield  {author} {\bibinfo {author} {\bibfnamefont {Y.}~\bibnamefont {Cao}}, \bibinfo {author} {\bibfnamefont {V.}~\bibnamefont {Fatemi}}, \bibinfo {author} {\bibfnamefont {S.}~\bibnamefont {Fang}}, \bibinfo {author} {\bibfnamefont {K.}~\bibnamefont {Watanabe}}, \bibinfo {author} {\bibfnamefont {T.}~\bibnamefont {Taniguchi}}, \bibinfo {author} {\bibfnamefont {E.}~\bibnamefont {Kaxiras}},\ and\ \bibinfo {author} {\bibfnamefont {P.}~\bibnamefont {{Jarillo-Herrero}}},\ }\bibfield  {title} {\bibinfo {title} {Unconventional superconductivity in magic-angle graphene superlattices},\ }\href {https://doi.org/10.1038/nature26160} {\bibfield  {journal} {\bibinfo  {journal} {Nature}\ }\textbf {\bibinfo {volume} {556}},\ \bibinfo {pages} {43} (\bibinfo {year} {2018}{\natexlab{b}})}\BibitemShut {NoStop}%
\bibitem [{\citenamefont {Lu}\ \emph {et~al.}(2019)\citenamefont {Lu}, \citenamefont {Stepanov}, \citenamefont {Yang}, \citenamefont {Xie}, \citenamefont {Aamir}, \citenamefont {Das}, \citenamefont {Urgell}, \citenamefont {Watanabe}, \citenamefont {Taniguchi}, \citenamefont {Zhang}, \citenamefont {Bachtold}, \citenamefont {MacDonald},\ and\ \citenamefont {Efetov}}]{lu2019}%
  \BibitemOpen
  \bibfield  {author} {\bibinfo {author} {\bibfnamefont {X.}~\bibnamefont {Lu}}, \bibinfo {author} {\bibfnamefont {P.}~\bibnamefont {Stepanov}}, \bibinfo {author} {\bibfnamefont {W.}~\bibnamefont {Yang}}, \bibinfo {author} {\bibfnamefont {M.}~\bibnamefont {Xie}}, \bibinfo {author} {\bibfnamefont {M.~A.}\ \bibnamefont {Aamir}}, \bibinfo {author} {\bibfnamefont {I.}~\bibnamefont {Das}}, \bibinfo {author} {\bibfnamefont {C.}~\bibnamefont {Urgell}}, \bibinfo {author} {\bibfnamefont {K.}~\bibnamefont {Watanabe}}, \bibinfo {author} {\bibfnamefont {T.}~\bibnamefont {Taniguchi}}, \bibinfo {author} {\bibfnamefont {G.}~\bibnamefont {Zhang}}, \bibinfo {author} {\bibfnamefont {A.}~\bibnamefont {Bachtold}}, \bibinfo {author} {\bibfnamefont {A.~H.}\ \bibnamefont {MacDonald}},\ and\ \bibinfo {author} {\bibfnamefont {D.~K.}\ \bibnamefont {Efetov}},\ }\bibfield  {title} {\bibinfo {title} {Superconductors, orbital magnets and correlated states in magic-angle bilayer graphene},\ }\href
  {https://doi.org/10.1038/s41586-019-1695-0} {\bibfield  {journal} {\bibinfo  {journal} {Nature}\ }\textbf {\bibinfo {volume} {574}},\ \bibinfo {pages} {653} (\bibinfo {year} {2019})}\BibitemShut {NoStop}%
\bibitem [{\citenamefont {Saito}\ \emph {et~al.}(2020)\citenamefont {Saito}, \citenamefont {Ge}, \citenamefont {Watanabe}, \citenamefont {Taniguchi},\ and\ \citenamefont {Young}}]{saito2020}%
  \BibitemOpen
  \bibfield  {author} {\bibinfo {author} {\bibfnamefont {Y.}~\bibnamefont {Saito}}, \bibinfo {author} {\bibfnamefont {J.}~\bibnamefont {Ge}}, \bibinfo {author} {\bibfnamefont {K.}~\bibnamefont {Watanabe}}, \bibinfo {author} {\bibfnamefont {T.}~\bibnamefont {Taniguchi}},\ and\ \bibinfo {author} {\bibfnamefont {A.~F.}\ \bibnamefont {Young}},\ }\bibfield  {title} {\bibinfo {title} {Independent superconductors and correlated insulators in twisted bilayer graphene},\ }\href {https://doi.org/10.1038/s41567-020-0928-3} {\bibfield  {journal} {\bibinfo  {journal} {Nat. Phys.}\ }\textbf {\bibinfo {volume} {16}},\ \bibinfo {pages} {926} (\bibinfo {year} {2020})}\BibitemShut {NoStop}%
\bibitem [{\citenamefont {Stepanov}\ \emph {et~al.}(2020)\citenamefont {Stepanov}, \citenamefont {Das}, \citenamefont {Lu}, \citenamefont {Fahimniya}, \citenamefont {Watanabe}, \citenamefont {Taniguchi}, \citenamefont {Koppens}, \citenamefont {Lischner}, \citenamefont {Levitov},\ and\ \citenamefont {Efetov}}]{stepanov2020}%
  \BibitemOpen
  \bibfield  {author} {\bibinfo {author} {\bibfnamefont {P.}~\bibnamefont {Stepanov}}, \bibinfo {author} {\bibfnamefont {I.}~\bibnamefont {Das}}, \bibinfo {author} {\bibfnamefont {X.}~\bibnamefont {Lu}}, \bibinfo {author} {\bibfnamefont {A.}~\bibnamefont {Fahimniya}}, \bibinfo {author} {\bibfnamefont {K.}~\bibnamefont {Watanabe}}, \bibinfo {author} {\bibfnamefont {T.}~\bibnamefont {Taniguchi}}, \bibinfo {author} {\bibfnamefont {F.~H.~L.}\ \bibnamefont {Koppens}}, \bibinfo {author} {\bibfnamefont {J.}~\bibnamefont {Lischner}}, \bibinfo {author} {\bibfnamefont {L.}~\bibnamefont {Levitov}},\ and\ \bibinfo {author} {\bibfnamefont {D.~K.}\ \bibnamefont {Efetov}},\ }\bibfield  {title} {\bibinfo {title} {Untying the insulating and superconducting orders in magic-angle graphene},\ }\href {https://doi.org/10.1038/s41586-020-2459-6} {\bibfield  {journal} {\bibinfo  {journal} {Nature}\ }\textbf {\bibinfo {volume} {583}},\ \bibinfo {pages} {375} (\bibinfo {year} {2020})}\BibitemShut {NoStop}%
\bibitem [{\citenamefont {Cao}\ \emph {et~al.}(2021)\citenamefont {Cao}, \citenamefont {{Rodan-Legrain}}, \citenamefont {Park}, \citenamefont {Yuan}, \citenamefont {Watanabe}, \citenamefont {Taniguchi}, \citenamefont {Fernandes}, \citenamefont {Fu},\ and\ \citenamefont {{Jarillo-Herrero}}}]{cao2021a}%
  \BibitemOpen
  \bibfield  {author} {\bibinfo {author} {\bibfnamefont {Y.}~\bibnamefont {Cao}}, \bibinfo {author} {\bibfnamefont {D.}~\bibnamefont {{Rodan-Legrain}}}, \bibinfo {author} {\bibfnamefont {J.~M.}\ \bibnamefont {Park}}, \bibinfo {author} {\bibfnamefont {N.~F.~Q.}\ \bibnamefont {Yuan}}, \bibinfo {author} {\bibfnamefont {K.}~\bibnamefont {Watanabe}}, \bibinfo {author} {\bibfnamefont {T.}~\bibnamefont {Taniguchi}}, \bibinfo {author} {\bibfnamefont {R.~M.}\ \bibnamefont {Fernandes}}, \bibinfo {author} {\bibfnamefont {L.}~\bibnamefont {Fu}},\ and\ \bibinfo {author} {\bibfnamefont {P.}~\bibnamefont {{Jarillo-Herrero}}},\ }\bibfield  {title} {\bibinfo {title} {Nematicity and competing orders in superconducting magic-angle graphene},\ }\href {https://doi.org/10.1126/science.abc2836} {\bibfield  {journal} {\bibinfo  {journal} {Science}\ }\textbf {\bibinfo {volume} {372}},\ \bibinfo {pages} {264} (\bibinfo {year} {2021})}\BibitemShut {NoStop}%
\bibitem [{\citenamefont {Liu}\ \emph {et~al.}(2021)\citenamefont {Liu}, \citenamefont {Wang}, \citenamefont {Watanabe}, \citenamefont {Taniguchi}, \citenamefont {Vafek},\ and\ \citenamefont {Li}}]{liu2021a}%
  \BibitemOpen
  \bibfield  {author} {\bibinfo {author} {\bibfnamefont {X.}~\bibnamefont {Liu}}, \bibinfo {author} {\bibfnamefont {Z.}~\bibnamefont {Wang}}, \bibinfo {author} {\bibfnamefont {K.}~\bibnamefont {Watanabe}}, \bibinfo {author} {\bibfnamefont {T.}~\bibnamefont {Taniguchi}}, \bibinfo {author} {\bibfnamefont {O.}~\bibnamefont {Vafek}},\ and\ \bibinfo {author} {\bibfnamefont {J.~I.~A.}\ \bibnamefont {Li}},\ }\bibfield  {title} {\bibinfo {title} {Tuning electron correlation in magic-angle twisted bilayer graphene using {{Coulomb}} screening},\ }\href {https://doi.org/10.1126/science.abb8754} {\bibfield  {journal} {\bibinfo  {journal} {Science}\ }\textbf {\bibinfo {volume} {371}},\ \bibinfo {pages} {1261} (\bibinfo {year} {2021})}\BibitemShut {NoStop}%
\bibitem [{\citenamefont {Sharpe}\ \emph {et~al.}(2019)\citenamefont {Sharpe}, \citenamefont {Fox}, \citenamefont {Barnard}, \citenamefont {Finney}, \citenamefont {Watanabe}, \citenamefont {Taniguchi}, \citenamefont {Kastner},\ and\ \citenamefont {{Goldhaber-Gordon}}}]{sharpe2019}%
  \BibitemOpen
  \bibfield  {author} {\bibinfo {author} {\bibfnamefont {A.~L.}\ \bibnamefont {Sharpe}}, \bibinfo {author} {\bibfnamefont {E.~J.}\ \bibnamefont {Fox}}, \bibinfo {author} {\bibfnamefont {A.~W.}\ \bibnamefont {Barnard}}, \bibinfo {author} {\bibfnamefont {J.}~\bibnamefont {Finney}}, \bibinfo {author} {\bibfnamefont {K.}~\bibnamefont {Watanabe}}, \bibinfo {author} {\bibfnamefont {T.}~\bibnamefont {Taniguchi}}, \bibinfo {author} {\bibfnamefont {M.~A.}\ \bibnamefont {Kastner}},\ and\ \bibinfo {author} {\bibfnamefont {D.}~\bibnamefont {{Goldhaber-Gordon}}},\ }\bibfield  {title} {\bibinfo {title} {Emergent ferromagnetism near three-quarters filling in twisted bilayer graphene},\ }\href {https://doi.org/10.1126/science.aaw3780} {\bibfield  {journal} {\bibinfo  {journal} {Science}\ }\textbf {\bibinfo {volume} {365}},\ \bibinfo {pages} {605} (\bibinfo {year} {2019})}\BibitemShut {NoStop}%
\bibitem [{\citenamefont {Serlin}\ \emph {et~al.}(2020)\citenamefont {Serlin}, \citenamefont {Tschirhart}, \citenamefont {Polshyn}, \citenamefont {Zhang}, \citenamefont {Zhu}, \citenamefont {Watanabe}, \citenamefont {Taniguchi}, \citenamefont {Balents},\ and\ \citenamefont {Young}}]{serlin2020}%
  \BibitemOpen
  \bibfield  {author} {\bibinfo {author} {\bibfnamefont {M.}~\bibnamefont {Serlin}}, \bibinfo {author} {\bibfnamefont {C.~L.}\ \bibnamefont {Tschirhart}}, \bibinfo {author} {\bibfnamefont {H.}~\bibnamefont {Polshyn}}, \bibinfo {author} {\bibfnamefont {Y.}~\bibnamefont {Zhang}}, \bibinfo {author} {\bibfnamefont {J.}~\bibnamefont {Zhu}}, \bibinfo {author} {\bibfnamefont {K.}~\bibnamefont {Watanabe}}, \bibinfo {author} {\bibfnamefont {T.}~\bibnamefont {Taniguchi}}, \bibinfo {author} {\bibfnamefont {L.}~\bibnamefont {Balents}},\ and\ \bibinfo {author} {\bibfnamefont {A.~F.}\ \bibnamefont {Young}},\ }\bibfield  {title} {\bibinfo {title} {Intrinsic quantized anomalous {{Hall}} effect in a moir\'e heterostructure},\ }\href {https://doi.org/10.1126/science.aay5533} {\bibfield  {journal} {\bibinfo  {journal} {Science}\ }\textbf {\bibinfo {volume} {367}},\ \bibinfo {pages} {900} (\bibinfo {year} {2020})}\BibitemShut {NoStop}%
\bibitem [{\citenamefont {Li}\ \emph {et~al.}(2010)\citenamefont {Li}, \citenamefont {Luican}, \citenamefont {Lopes~dos Santos}, \citenamefont {Castro~Neto}, \citenamefont {Reina}, \citenamefont {Kong},\ and\ \citenamefont {Andrei}}]{Guohong2010}%
  \BibitemOpen
  \bibfield  {author} {\bibinfo {author} {\bibfnamefont {G.}~\bibnamefont {Li}}, \bibinfo {author} {\bibfnamefont {A.}~\bibnamefont {Luican}}, \bibinfo {author} {\bibfnamefont {J.~M.~B.}\ \bibnamefont {Lopes~dos Santos}}, \bibinfo {author} {\bibfnamefont {A.~H.}\ \bibnamefont {Castro~Neto}}, \bibinfo {author} {\bibfnamefont {A.}~\bibnamefont {Reina}}, \bibinfo {author} {\bibfnamefont {J.}~\bibnamefont {Kong}},\ and\ \bibinfo {author} {\bibfnamefont {E.~Y.}\ \bibnamefont {Andrei}},\ }\bibfield  {title} {\bibinfo {title} {Observation of van hove singularities in twisted graphene layers},\ }\href {https://doi.org/10.1038/nphys1463} {\bibfield  {journal} {\bibinfo  {journal} {Nature Physics}\ }\textbf {\bibinfo {volume} {6}},\ \bibinfo {pages} {109} (\bibinfo {year} {2010})}\BibitemShut {NoStop}%
\bibitem [{\citenamefont {Brihuega}\ \emph {et~al.}(2012)\citenamefont {Brihuega}, \citenamefont {Mallet}, \citenamefont {Gonz\'alez-Herrero}, \citenamefont {Trambly~de Laissardi\`ere}, \citenamefont {Ugeda}, \citenamefont {Magaud}, \citenamefont {G\'omez-Rodr\'{\i}guez}, \citenamefont {Yndur\'ain},\ and\ \citenamefont {Veuillen}}]{PhysRevLett.109.196802}%
  \BibitemOpen
  \bibfield  {author} {\bibinfo {author} {\bibfnamefont {I.}~\bibnamefont {Brihuega}}, \bibinfo {author} {\bibfnamefont {P.}~\bibnamefont {Mallet}}, \bibinfo {author} {\bibfnamefont {H.}~\bibnamefont {Gonz\'alez-Herrero}}, \bibinfo {author} {\bibfnamefont {G.}~\bibnamefont {Trambly~de Laissardi\`ere}}, \bibinfo {author} {\bibfnamefont {M.~M.}\ \bibnamefont {Ugeda}}, \bibinfo {author} {\bibfnamefont {L.}~\bibnamefont {Magaud}}, \bibinfo {author} {\bibfnamefont {J.~M.}\ \bibnamefont {G\'omez-Rodr\'{\i}guez}}, \bibinfo {author} {\bibfnamefont {F.}~\bibnamefont {Yndur\'ain}},\ and\ \bibinfo {author} {\bibfnamefont {J.-Y.}\ \bibnamefont {Veuillen}},\ }\bibfield  {title} {\bibinfo {title} {Unraveling the intrinsic and robust nature of van hove singularities in twisted bilayer graphene by scanning tunneling microscopy and theoretical analysis},\ }\href {https://doi.org/10.1103/PhysRevLett.109.196802} {\bibfield  {journal} {\bibinfo  {journal} {Phys. Rev. Lett.}\ }\textbf {\bibinfo {volume} {109}},\ \bibinfo {pages}
  {196802} (\bibinfo {year} {2012})}\BibitemShut {NoStop}%
\bibitem [{\citenamefont {Kerelsky}\ \emph {et~al.}(2019)\citenamefont {Kerelsky}, \citenamefont {McGilly}, \citenamefont {Kennes}, \citenamefont {Xian}, \citenamefont {Yankowitz}, \citenamefont {Chen}, \citenamefont {Watanabe}, \citenamefont {Taniguchi}, \citenamefont {Hone}, \citenamefont {Dean}, \citenamefont {Rubio},\ and\ \citenamefont {Pasupathy}}]{Kerelsky2019}%
  \BibitemOpen
  \bibfield  {author} {\bibinfo {author} {\bibfnamefont {A.}~\bibnamefont {Kerelsky}}, \bibinfo {author} {\bibfnamefont {L.~J.}\ \bibnamefont {McGilly}}, \bibinfo {author} {\bibfnamefont {D.~M.}\ \bibnamefont {Kennes}}, \bibinfo {author} {\bibfnamefont {L.}~\bibnamefont {Xian}}, \bibinfo {author} {\bibfnamefont {M.}~\bibnamefont {Yankowitz}}, \bibinfo {author} {\bibfnamefont {S.}~\bibnamefont {Chen}}, \bibinfo {author} {\bibfnamefont {K.}~\bibnamefont {Watanabe}}, \bibinfo {author} {\bibfnamefont {T.}~\bibnamefont {Taniguchi}}, \bibinfo {author} {\bibfnamefont {J.}~\bibnamefont {Hone}}, \bibinfo {author} {\bibfnamefont {C.}~\bibnamefont {Dean}}, \bibinfo {author} {\bibfnamefont {A.}~\bibnamefont {Rubio}},\ and\ \bibinfo {author} {\bibfnamefont {A.~N.}\ \bibnamefont {Pasupathy}},\ }\bibfield  {title} {\bibinfo {title} {Maximized electron interactions at the magic angle in twisted bilayer graphene},\ }\href {https://doi.org/10.1038/s41586-019-1431-9} {\bibfield  {journal} {\bibinfo  {journal} {Nature}\ }\textbf
  {\bibinfo {volume} {572}},\ \bibinfo {pages} {95} (\bibinfo {year} {2019})}\BibitemShut {NoStop}%
\bibitem [{\citenamefont {Utama}\ \emph {et~al.}(2021)\citenamefont {Utama}, \citenamefont {Koch}, \citenamefont {Lee}, \citenamefont {Leconte}, \citenamefont {Li}, \citenamefont {Zhao}, \citenamefont {Jiang}, \citenamefont {Zhu}, \citenamefont {Watanabe}, \citenamefont {Taniguchi}, \citenamefont {Ashby}, \citenamefont {{Weber-Bargioni}}, \citenamefont {Zettl}, \citenamefont {Jozwiak}, \citenamefont {Jung}, \citenamefont {Rotenberg}, \citenamefont {Bostwick},\ and\ \citenamefont {Wang}}]{utama2021}%
  \BibitemOpen
  \bibfield  {author} {\bibinfo {author} {\bibfnamefont {M.~I.~B.}\ \bibnamefont {Utama}}, \bibinfo {author} {\bibfnamefont {R.~J.}\ \bibnamefont {Koch}}, \bibinfo {author} {\bibfnamefont {K.}~\bibnamefont {Lee}}, \bibinfo {author} {\bibfnamefont {N.}~\bibnamefont {Leconte}}, \bibinfo {author} {\bibfnamefont {H.}~\bibnamefont {Li}}, \bibinfo {author} {\bibfnamefont {S.}~\bibnamefont {Zhao}}, \bibinfo {author} {\bibfnamefont {L.}~\bibnamefont {Jiang}}, \bibinfo {author} {\bibfnamefont {J.}~\bibnamefont {Zhu}}, \bibinfo {author} {\bibfnamefont {K.}~\bibnamefont {Watanabe}}, \bibinfo {author} {\bibfnamefont {T.}~\bibnamefont {Taniguchi}}, \bibinfo {author} {\bibfnamefont {P.~D.}\ \bibnamefont {Ashby}}, \bibinfo {author} {\bibfnamefont {A.}~\bibnamefont {{Weber-Bargioni}}}, \bibinfo {author} {\bibfnamefont {A.}~\bibnamefont {Zettl}}, \bibinfo {author} {\bibfnamefont {C.}~\bibnamefont {Jozwiak}}, \bibinfo {author} {\bibfnamefont {J.}~\bibnamefont {Jung}}, \bibinfo {author} {\bibfnamefont {E.}~\bibnamefont
  {Rotenberg}}, \bibinfo {author} {\bibfnamefont {A.}~\bibnamefont {Bostwick}},\ and\ \bibinfo {author} {\bibfnamefont {F.}~\bibnamefont {Wang}},\ }\bibfield  {title} {\bibinfo {title} {Visualization of the flat electronic band in twisted bilayer graphene near the magic angle twist},\ }\href {https://doi.org/10.1038/s41567-020-0974-x} {\bibfield  {journal} {\bibinfo  {journal} {Nat. Phys.}\ }\textbf {\bibinfo {volume} {17}},\ \bibinfo {pages} {184} (\bibinfo {year} {2021})}\BibitemShut {NoStop}%
\bibitem [{\citenamefont {Lisi}\ \emph {et~al.}(2021)\citenamefont {Lisi}, \citenamefont {Lu}, \citenamefont {Benschop}, \citenamefont {{de Jong}}, \citenamefont {Stepanov}, \citenamefont {Duran}, \citenamefont {Margot}, \citenamefont {Cucchi}, \citenamefont {Cappelli}, \citenamefont {Hunter}, \citenamefont {Tamai}, \citenamefont {Kandyba}, \citenamefont {Giampietri}, \citenamefont {Barinov}, \citenamefont {Jobst}, \citenamefont {Stalman}, \citenamefont {Leeuwenhoek}, \citenamefont {Watanabe}, \citenamefont {Taniguchi}, \citenamefont {Rademaker}, \citenamefont {{van der Molen}}, \citenamefont {Allan}, \citenamefont {Efetov},\ and\ \citenamefont {Baumberger}}]{Lisi2021}%
  \BibitemOpen
  \bibfield  {author} {\bibinfo {author} {\bibfnamefont {S.}~\bibnamefont {Lisi}}, \bibinfo {author} {\bibfnamefont {X.}~\bibnamefont {Lu}}, \bibinfo {author} {\bibfnamefont {T.}~\bibnamefont {Benschop}}, \bibinfo {author} {\bibfnamefont {T.~A.}\ \bibnamefont {{de Jong}}}, \bibinfo {author} {\bibfnamefont {P.}~\bibnamefont {Stepanov}}, \bibinfo {author} {\bibfnamefont {J.~R.}\ \bibnamefont {Duran}}, \bibinfo {author} {\bibfnamefont {F.}~\bibnamefont {Margot}}, \bibinfo {author} {\bibfnamefont {I.}~\bibnamefont {Cucchi}}, \bibinfo {author} {\bibfnamefont {E.}~\bibnamefont {Cappelli}}, \bibinfo {author} {\bibfnamefont {A.}~\bibnamefont {Hunter}}, \bibinfo {author} {\bibfnamefont {A.}~\bibnamefont {Tamai}}, \bibinfo {author} {\bibfnamefont {V.}~\bibnamefont {Kandyba}}, \bibinfo {author} {\bibfnamefont {A.}~\bibnamefont {Giampietri}}, \bibinfo {author} {\bibfnamefont {A.}~\bibnamefont {Barinov}}, \bibinfo {author} {\bibfnamefont {J.}~\bibnamefont {Jobst}}, \bibinfo {author} {\bibfnamefont {V.}~\bibnamefont
  {Stalman}}, \bibinfo {author} {\bibfnamefont {M.}~\bibnamefont {Leeuwenhoek}}, \bibinfo {author} {\bibfnamefont {K.}~\bibnamefont {Watanabe}}, \bibinfo {author} {\bibfnamefont {T.}~\bibnamefont {Taniguchi}}, \bibinfo {author} {\bibfnamefont {L.}~\bibnamefont {Rademaker}}, \bibinfo {author} {\bibfnamefont {S.~J.}\ \bibnamefont {{van der Molen}}}, \bibinfo {author} {\bibfnamefont {M.~P.}\ \bibnamefont {Allan}}, \bibinfo {author} {\bibfnamefont {D.~K.}\ \bibnamefont {Efetov}},\ and\ \bibinfo {author} {\bibfnamefont {F.}~\bibnamefont {Baumberger}},\ }\bibfield  {title} {\bibinfo {title} {Observation of flat bands in twisted bilayer graphene},\ }\href {https://doi.org/10.1038/s41567-020-01041-x} {\bibfield  {journal} {\bibinfo  {journal} {Nat. Phys.}\ }\textbf {\bibinfo {volume} {17}},\ \bibinfo {pages} {189} (\bibinfo {year} {2021})}\BibitemShut {NoStop}%
\bibitem [{\citenamefont {Nuckolls}\ \emph {et~al.}(2020)\citenamefont {Nuckolls}, \citenamefont {Oh}, \citenamefont {Wong}, \citenamefont {Lian}, \citenamefont {Watanabe}, \citenamefont {Taniguchi}, \citenamefont {Bernevig},\ and\ \citenamefont {Yazdani}}]{nuckolls2020}%
  \BibitemOpen
  \bibfield  {author} {\bibinfo {author} {\bibfnamefont {K.~P.}\ \bibnamefont {Nuckolls}}, \bibinfo {author} {\bibfnamefont {M.}~\bibnamefont {Oh}}, \bibinfo {author} {\bibfnamefont {D.}~\bibnamefont {Wong}}, \bibinfo {author} {\bibfnamefont {B.}~\bibnamefont {Lian}}, \bibinfo {author} {\bibfnamefont {K.}~\bibnamefont {Watanabe}}, \bibinfo {author} {\bibfnamefont {T.}~\bibnamefont {Taniguchi}}, \bibinfo {author} {\bibfnamefont {B.~A.}\ \bibnamefont {Bernevig}},\ and\ \bibinfo {author} {\bibfnamefont {A.}~\bibnamefont {Yazdani}},\ }\bibfield  {title} {\bibinfo {title} {Strongly correlated {{Chern}} insulators in magic-angle twisted bilayer graphene},\ }\href {https://doi.org/10.1038/s41586-020-3028-8} {\bibfield  {journal} {\bibinfo  {journal} {Nature}\ ,\ \bibinfo {pages} {1}} (\bibinfo {year} {2020})}\BibitemShut {NoStop}%
\bibitem [{\citenamefont {Choi}\ \emph {et~al.}(2021)\citenamefont {Choi}, \citenamefont {Kim}, \citenamefont {Peng}, \citenamefont {Thomson}, \citenamefont {Lewandowski}, \citenamefont {Polski}, \citenamefont {Zhang}, \citenamefont {Arora}, \citenamefont {Watanabe}, \citenamefont {Taniguchi}, \citenamefont {Alicea},\ and\ \citenamefont {{Nadj-Perge}}}]{choi2021a}%
  \BibitemOpen
  \bibfield  {author} {\bibinfo {author} {\bibfnamefont {Y.}~\bibnamefont {Choi}}, \bibinfo {author} {\bibfnamefont {H.}~\bibnamefont {Kim}}, \bibinfo {author} {\bibfnamefont {Y.}~\bibnamefont {Peng}}, \bibinfo {author} {\bibfnamefont {A.}~\bibnamefont {Thomson}}, \bibinfo {author} {\bibfnamefont {C.}~\bibnamefont {Lewandowski}}, \bibinfo {author} {\bibfnamefont {R.}~\bibnamefont {Polski}}, \bibinfo {author} {\bibfnamefont {Y.}~\bibnamefont {Zhang}}, \bibinfo {author} {\bibfnamefont {H.~S.}\ \bibnamefont {Arora}}, \bibinfo {author} {\bibfnamefont {K.}~\bibnamefont {Watanabe}}, \bibinfo {author} {\bibfnamefont {T.}~\bibnamefont {Taniguchi}}, \bibinfo {author} {\bibfnamefont {J.}~\bibnamefont {Alicea}},\ and\ \bibinfo {author} {\bibfnamefont {S.}~\bibnamefont {{Nadj-Perge}}},\ }\bibfield  {title} {\bibinfo {title} {Correlation-driven topological phases in magic-angle twisted bilayer graphene},\ }\href {https://doi.org/10.1038/s41586-020-03159-7} {\bibfield  {journal} {\bibinfo  {journal} {Nature}\ }\textbf
  {\bibinfo {volume} {589}},\ \bibinfo {pages} {536} (\bibinfo {year} {2021})}\BibitemShut {NoStop}%
\bibitem [{\citenamefont {Wu}\ \emph {et~al.}(2021)\citenamefont {Wu}, \citenamefont {Zhang}, \citenamefont {Watanabe}, \citenamefont {Taniguchi},\ and\ \citenamefont {Andrei}}]{wu2021}%
  \BibitemOpen
  \bibfield  {author} {\bibinfo {author} {\bibfnamefont {S.}~\bibnamefont {Wu}}, \bibinfo {author} {\bibfnamefont {Z.}~\bibnamefont {Zhang}}, \bibinfo {author} {\bibfnamefont {K.}~\bibnamefont {Watanabe}}, \bibinfo {author} {\bibfnamefont {T.}~\bibnamefont {Taniguchi}},\ and\ \bibinfo {author} {\bibfnamefont {E.~Y.}\ \bibnamefont {Andrei}},\ }\bibfield  {title} {\bibinfo {title} {Chern insulators, van {{Hove}} singularities and topological flat bands in magic-angle twisted bilayer graphene},\ }\href {https://doi.org/10.1038/s41563-020-00911-2} {\bibfield  {journal} {\bibinfo  {journal} {Nat. Mater.}\ }\textbf {\bibinfo {volume} {20}},\ \bibinfo {pages} {488} (\bibinfo {year} {2021})}\BibitemShut {NoStop}%
\bibitem [{\citenamefont {Saito}\ \emph {et~al.}(2021)\citenamefont {Saito}, \citenamefont {Ge}, \citenamefont {Rademaker}, \citenamefont {Watanabe}, \citenamefont {Taniguchi}, \citenamefont {Abanin},\ and\ \citenamefont {Young}}]{saito2021a}%
  \BibitemOpen
  \bibfield  {author} {\bibinfo {author} {\bibfnamefont {Y.}~\bibnamefont {Saito}}, \bibinfo {author} {\bibfnamefont {J.}~\bibnamefont {Ge}}, \bibinfo {author} {\bibfnamefont {L.}~\bibnamefont {Rademaker}}, \bibinfo {author} {\bibfnamefont {K.}~\bibnamefont {Watanabe}}, \bibinfo {author} {\bibfnamefont {T.}~\bibnamefont {Taniguchi}}, \bibinfo {author} {\bibfnamefont {D.~A.}\ \bibnamefont {Abanin}},\ and\ \bibinfo {author} {\bibfnamefont {A.~F.}\ \bibnamefont {Young}},\ }\bibfield  {title} {\bibinfo {title} {Hofstadter subband ferromagnetism and symmetry-broken {{Chern}} insulators in twisted bilayer graphene},\ }\href {https://doi.org/10.1038/s41567-020-01129-4} {\bibfield  {journal} {\bibinfo  {journal} {Nat. Phys.}\ }\textbf {\bibinfo {volume} {17}},\ \bibinfo {pages} {478} (\bibinfo {year} {2021})}\BibitemShut {NoStop}%
\bibitem [{\citenamefont {Das}\ \emph {et~al.}(2021)\citenamefont {Das}, \citenamefont {Lu}, \citenamefont {{Herzog-Arbeitman}}, \citenamefont {Song}, \citenamefont {Watanabe}, \citenamefont {Taniguchi}, \citenamefont {Bernevig},\ and\ \citenamefont {Efetov}}]{das2021a}%
  \BibitemOpen
  \bibfield  {author} {\bibinfo {author} {\bibfnamefont {I.}~\bibnamefont {Das}}, \bibinfo {author} {\bibfnamefont {X.}~\bibnamefont {Lu}}, \bibinfo {author} {\bibfnamefont {J.}~\bibnamefont {{Herzog-Arbeitman}}}, \bibinfo {author} {\bibfnamefont {Z.-D.}\ \bibnamefont {Song}}, \bibinfo {author} {\bibfnamefont {K.}~\bibnamefont {Watanabe}}, \bibinfo {author} {\bibfnamefont {T.}~\bibnamefont {Taniguchi}}, \bibinfo {author} {\bibfnamefont {B.~A.}\ \bibnamefont {Bernevig}},\ and\ \bibinfo {author} {\bibfnamefont {D.~K.}\ \bibnamefont {Efetov}},\ }\bibfield  {title} {\bibinfo {title} {Symmetry-broken {{Chern}} insulators and {{Rashba-like Landau-level}} crossings in magic-angle bilayer graphene},\ }\href {https://doi.org/10.1038/s41567-021-01186-3} {\bibfield  {journal} {\bibinfo  {journal} {Nat. Phys.}\ }\textbf {\bibinfo {volume} {17}},\ \bibinfo {pages} {710} (\bibinfo {year} {2021})}\BibitemShut {NoStop}%
\bibitem [{\citenamefont {Park}\ \emph {et~al.}(2021)\citenamefont {Park}, \citenamefont {Cao}, \citenamefont {Watanabe}, \citenamefont {Taniguchi},\ and\ \citenamefont {{Jarillo-Herrero}}}]{park2021a}%
  \BibitemOpen
  \bibfield  {author} {\bibinfo {author} {\bibfnamefont {J.~M.}\ \bibnamefont {Park}}, \bibinfo {author} {\bibfnamefont {Y.}~\bibnamefont {Cao}}, \bibinfo {author} {\bibfnamefont {K.}~\bibnamefont {Watanabe}}, \bibinfo {author} {\bibfnamefont {T.}~\bibnamefont {Taniguchi}},\ and\ \bibinfo {author} {\bibfnamefont {P.}~\bibnamefont {{Jarillo-Herrero}}},\ }\bibfield  {title} {\bibinfo {title} {Flavour {{Hund}}'s coupling, {{Chern}} gaps and charge diffusivity in moir{\'e} graphene},\ }\href {https://doi.org/10.1038/s41586-021-03366-w} {\bibfield  {journal} {\bibinfo  {journal} {Nature}\ }\textbf {\bibinfo {volume} {592}},\ \bibinfo {pages} {43} (\bibinfo {year} {2021})}\BibitemShut {NoStop}%
\bibitem [{\citenamefont {Pierce}\ \emph {et~al.}(2021)\citenamefont {Pierce}, \citenamefont {Xie}, \citenamefont {Park}, \citenamefont {Khalaf}, \citenamefont {Lee}, \citenamefont {Cao}, \citenamefont {Parker}, \citenamefont {Forrester}, \citenamefont {Chen}, \citenamefont {Watanabe}, \citenamefont {Taniguchi}, \citenamefont {Vishwanath}, \citenamefont {{Jarillo-Herrero}},\ and\ \citenamefont {Yacoby}}]{pierce2021}%
  \BibitemOpen
  \bibfield  {author} {\bibinfo {author} {\bibfnamefont {A.~T.}\ \bibnamefont {Pierce}}, \bibinfo {author} {\bibfnamefont {Y.}~\bibnamefont {Xie}}, \bibinfo {author} {\bibfnamefont {J.~M.}\ \bibnamefont {Park}}, \bibinfo {author} {\bibfnamefont {E.}~\bibnamefont {Khalaf}}, \bibinfo {author} {\bibfnamefont {S.~H.}\ \bibnamefont {Lee}}, \bibinfo {author} {\bibfnamefont {Y.}~\bibnamefont {Cao}}, \bibinfo {author} {\bibfnamefont {D.~E.}\ \bibnamefont {Parker}}, \bibinfo {author} {\bibfnamefont {P.~R.}\ \bibnamefont {Forrester}}, \bibinfo {author} {\bibfnamefont {S.}~\bibnamefont {Chen}}, \bibinfo {author} {\bibfnamefont {K.}~\bibnamefont {Watanabe}}, \bibinfo {author} {\bibfnamefont {T.}~\bibnamefont {Taniguchi}}, \bibinfo {author} {\bibfnamefont {A.}~\bibnamefont {Vishwanath}}, \bibinfo {author} {\bibfnamefont {P.}~\bibnamefont {{Jarillo-Herrero}}},\ and\ \bibinfo {author} {\bibfnamefont {A.}~\bibnamefont {Yacoby}},\ }\bibfield  {title} {\bibinfo {title} {Unconventional sequence of correlated {{Chern}}
  insulators in magic-angle twisted bilayer graphene},\ }\href {https://doi.org/10.1038/s41567-021-01347-4} {\bibfield  {journal} {\bibinfo  {journal} {Nat. Phys.}\ }\textbf {\bibinfo {volume} {17}},\ \bibinfo {pages} {1210} (\bibinfo {year} {2021})}\BibitemShut {NoStop}%
\bibitem [{\citenamefont {Xie}\ \emph {et~al.}(2021)\citenamefont {Xie}, \citenamefont {Pierce}, \citenamefont {Park}, \citenamefont {Parker}, \citenamefont {Khalaf}, \citenamefont {Ledwith}, \citenamefont {Cao}, \citenamefont {Lee}, \citenamefont {Chen}, \citenamefont {Forrester}, \citenamefont {Watanabe}, \citenamefont {Taniguchi}, \citenamefont {Vishwanath}, \citenamefont {{Jarillo-Herrero}},\ and\ \citenamefont {Yacoby}}]{xie2021b}%
  \BibitemOpen
  \bibfield  {author} {\bibinfo {author} {\bibfnamefont {Y.}~\bibnamefont {Xie}}, \bibinfo {author} {\bibfnamefont {A.~T.}\ \bibnamefont {Pierce}}, \bibinfo {author} {\bibfnamefont {J.~M.}\ \bibnamefont {Park}}, \bibinfo {author} {\bibfnamefont {D.~E.}\ \bibnamefont {Parker}}, \bibinfo {author} {\bibfnamefont {E.}~\bibnamefont {Khalaf}}, \bibinfo {author} {\bibfnamefont {P.}~\bibnamefont {Ledwith}}, \bibinfo {author} {\bibfnamefont {Y.}~\bibnamefont {Cao}}, \bibinfo {author} {\bibfnamefont {S.~H.}\ \bibnamefont {Lee}}, \bibinfo {author} {\bibfnamefont {S.}~\bibnamefont {Chen}}, \bibinfo {author} {\bibfnamefont {P.~R.}\ \bibnamefont {Forrester}}, \bibinfo {author} {\bibfnamefont {K.}~\bibnamefont {Watanabe}}, \bibinfo {author} {\bibfnamefont {T.}~\bibnamefont {Taniguchi}}, \bibinfo {author} {\bibfnamefont {A.}~\bibnamefont {Vishwanath}}, \bibinfo {author} {\bibfnamefont {P.}~\bibnamefont {{Jarillo-Herrero}}},\ and\ \bibinfo {author} {\bibfnamefont {A.}~\bibnamefont {Yacoby}},\ }\bibfield  {title} {\bibinfo
  {title} {Fractional {{Chern}} insulators in magic-angle twisted bilayer graphene},\ }\href {https://doi.org/10.1038/s41586-021-04002-3} {\bibfield  {journal} {\bibinfo  {journal} {Nature}\ }\textbf {\bibinfo {volume} {600}},\ \bibinfo {pages} {439} (\bibinfo {year} {2021})}\BibitemShut {NoStop}%
\bibitem [{\citenamefont {Yuan}\ and\ \citenamefont {Fu}(2018)}]{yuan2018}%
  \BibitemOpen
  \bibfield  {author} {\bibinfo {author} {\bibfnamefont {N.~F.~Q.}\ \bibnamefont {Yuan}}\ and\ \bibinfo {author} {\bibfnamefont {L.}~\bibnamefont {Fu}},\ }\bibfield  {title} {\bibinfo {title} {Model for the metal-insulator transition in graphene superlattices and beyond},\ }\href {https://doi.org/10.1103/PhysRevB.98.045103} {\bibfield  {journal} {\bibinfo  {journal} {Phys. Rev. B}\ }\textbf {\bibinfo {volume} {98}},\ \bibinfo {pages} {045103} (\bibinfo {year} {2018})}\BibitemShut {NoStop}%
\bibitem [{\citenamefont {Kang}\ and\ \citenamefont {Vafek}(2018)}]{kang2018}%
  \BibitemOpen
  \bibfield  {author} {\bibinfo {author} {\bibfnamefont {J.}~\bibnamefont {Kang}}\ and\ \bibinfo {author} {\bibfnamefont {O.}~\bibnamefont {Vafek}},\ }\bibfield  {title} {\bibinfo {title} {Symmetry, {{Maximally Localized Wannier States}}, and a {{Low-Energy Model}} for {{Twisted Bilayer Graphene Narrow Bands}}},\ }\href {https://doi.org/10.1103/PhysRevX.8.031088} {\bibfield  {journal} {\bibinfo  {journal} {Phys. Rev. X}\ }\textbf {\bibinfo {volume} {8}},\ \bibinfo {pages} {031088} (\bibinfo {year} {2018})}\BibitemShut {NoStop}%
\bibitem [{\citenamefont {Koshino}\ \emph {et~al.}(2018)\citenamefont {Koshino}, \citenamefont {Yuan}, \citenamefont {Koretsune}, \citenamefont {Ochi}, \citenamefont {Kuroki},\ and\ \citenamefont {Fu}}]{koshino2018}%
  \BibitemOpen
  \bibfield  {author} {\bibinfo {author} {\bibfnamefont {M.}~\bibnamefont {Koshino}}, \bibinfo {author} {\bibfnamefont {N.~F.~Q.}\ \bibnamefont {Yuan}}, \bibinfo {author} {\bibfnamefont {T.}~\bibnamefont {Koretsune}}, \bibinfo {author} {\bibfnamefont {M.}~\bibnamefont {Ochi}}, \bibinfo {author} {\bibfnamefont {K.}~\bibnamefont {Kuroki}},\ and\ \bibinfo {author} {\bibfnamefont {L.}~\bibnamefont {Fu}},\ }\bibfield  {title} {\bibinfo {title} {Maximally {{Localized Wannier Orbitals}} and the {{Extended Hubbard Model}} for {{Twisted Bilayer Graphene}}},\ }\href {https://doi.org/10.1103/PhysRevX.8.031087} {\bibfield  {journal} {\bibinfo  {journal} {Phys. Rev. X}\ }\textbf {\bibinfo {volume} {8}},\ \bibinfo {pages} {031087} (\bibinfo {year} {2018})}\BibitemShut {NoStop}%
\bibitem [{\citenamefont {Po}\ \emph {et~al.}(2018)\citenamefont {Po}, \citenamefont {Zou}, \citenamefont {Vishwanath},\ and\ \citenamefont {Senthil}}]{po2018b}%
  \BibitemOpen
  \bibfield  {author} {\bibinfo {author} {\bibfnamefont {H.~C.}\ \bibnamefont {Po}}, \bibinfo {author} {\bibfnamefont {L.}~\bibnamefont {Zou}}, \bibinfo {author} {\bibfnamefont {A.}~\bibnamefont {Vishwanath}},\ and\ \bibinfo {author} {\bibfnamefont {T.}~\bibnamefont {Senthil}},\ }\bibfield  {title} {\bibinfo {title} {Origin of {{Mott Insulating Behavior}} and {{Superconductivity}} in {{Twisted Bilayer Graphene}}},\ }\href {https://doi.org/10.1103/PhysRevX.8.031089} {\bibfield  {journal} {\bibinfo  {journal} {Phys. Rev. X}\ }\textbf {\bibinfo {volume} {8}},\ \bibinfo {pages} {031089} (\bibinfo {year} {2018})}\BibitemShut {NoStop}%
\bibitem [{\citenamefont {Rademaker}\ and\ \citenamefont {Mellado}(2018)}]{rademaker2018}%
  \BibitemOpen
  \bibfield  {author} {\bibinfo {author} {\bibfnamefont {L.}~\bibnamefont {Rademaker}}\ and\ \bibinfo {author} {\bibfnamefont {P.}~\bibnamefont {Mellado}},\ }\bibfield  {title} {\bibinfo {title} {Charge-transfer insulation in twisted bilayer graphene},\ }\href {https://doi.org/10.1103/PhysRevB.98.235158} {\bibfield  {journal} {\bibinfo  {journal} {Phys. Rev. B}\ }\textbf {\bibinfo {volume} {98}},\ \bibinfo {pages} {235158} (\bibinfo {year} {2018})}\BibitemShut {NoStop}%
\bibitem [{\citenamefont {Xie}\ and\ \citenamefont {MacDonald}(2020)}]{xie2020a}%
  \BibitemOpen
  \bibfield  {author} {\bibinfo {author} {\bibfnamefont {M.}~\bibnamefont {Xie}}\ and\ \bibinfo {author} {\bibfnamefont {A.~H.}\ \bibnamefont {MacDonald}},\ }\bibfield  {title} {\bibinfo {title} {Nature of the {{Correlated Insulator States}} in {{Twisted Bilayer Graphene}}},\ }\href {https://doi.org/10.1103/PhysRevLett.124.097601} {\bibfield  {journal} {\bibinfo  {journal} {Phys. Rev. Lett.}\ }\textbf {\bibinfo {volume} {124}},\ \bibinfo {pages} {097601} (\bibinfo {year} {2020})}\BibitemShut {NoStop}%
\bibitem [{\citenamefont {Song}\ and\ \citenamefont {Bernevig}(2022)}]{PhysRevLett.129.047601}%
  \BibitemOpen
  \bibfield  {author} {\bibinfo {author} {\bibfnamefont {Z.-D.}\ \bibnamefont {Song}}\ and\ \bibinfo {author} {\bibfnamefont {B.~A.}\ \bibnamefont {Bernevig}},\ }\bibfield  {title} {\bibinfo {title} {Magic-angle twisted bilayer graphene as a topological heavy fermion problem},\ }\href {https://doi.org/10.1103/PhysRevLett.129.047601} {\bibfield  {journal} {\bibinfo  {journal} {Phys. Rev. Lett.}\ }\textbf {\bibinfo {volume} {129}},\ \bibinfo {pages} {047601} (\bibinfo {year} {2022})}\BibitemShut {NoStop}%
\bibitem [{\citenamefont {Fang}\ and\ \citenamefont {Kaxiras}(2016)}]{PhysRevB.93.235153}%
  \BibitemOpen
  \bibfield  {author} {\bibinfo {author} {\bibfnamefont {S.}~\bibnamefont {Fang}}\ and\ \bibinfo {author} {\bibfnamefont {E.}~\bibnamefont {Kaxiras}},\ }\bibfield  {title} {\bibinfo {title} {Electronic structure theory of weakly interacting bilayers},\ }\href {https://doi.org/10.1103/PhysRevB.93.235153} {\bibfield  {journal} {\bibinfo  {journal} {Phys. Rev. B}\ }\textbf {\bibinfo {volume} {93}},\ \bibinfo {pages} {235153} (\bibinfo {year} {2016})}\BibitemShut {NoStop}%
\bibitem [{\citenamefont {Carr}\ \emph {et~al.}(2018{\natexlab{a}})\citenamefont {Carr}, \citenamefont {Fang}, \citenamefont {Jarillo-Herrero},\ and\ \citenamefont {Kaxiras}}]{PhysRevB.98.085144}%
  \BibitemOpen
  \bibfield  {author} {\bibinfo {author} {\bibfnamefont {S.}~\bibnamefont {Carr}}, \bibinfo {author} {\bibfnamefont {S.}~\bibnamefont {Fang}}, \bibinfo {author} {\bibfnamefont {P.}~\bibnamefont {Jarillo-Herrero}},\ and\ \bibinfo {author} {\bibfnamefont {E.}~\bibnamefont {Kaxiras}},\ }\bibfield  {title} {\bibinfo {title} {Pressure dependence of the magic twist angle in graphene superlattices},\ }\href {https://doi.org/10.1103/PhysRevB.98.085144} {\bibfield  {journal} {\bibinfo  {journal} {Phys. Rev. B}\ }\textbf {\bibinfo {volume} {98}},\ \bibinfo {pages} {085144} (\bibinfo {year} {2018}{\natexlab{a}})}\BibitemShut {NoStop}%
\bibitem [{\citenamefont {Nam}\ and\ \citenamefont {Koshino}(2017)}]{PhysRevB.96.075311}%
  \BibitemOpen
  \bibfield  {author} {\bibinfo {author} {\bibfnamefont {N.~N.~T.}\ \bibnamefont {Nam}}\ and\ \bibinfo {author} {\bibfnamefont {M.}~\bibnamefont {Koshino}},\ }\bibfield  {title} {\bibinfo {title} {Lattice relaxation and energy band modulation in twisted bilayer graphene},\ }\href {https://doi.org/10.1103/PhysRevB.96.075311} {\bibfield  {journal} {\bibinfo  {journal} {Phys. Rev. B}\ }\textbf {\bibinfo {volume} {96}},\ \bibinfo {pages} {075311} (\bibinfo {year} {2017})}\BibitemShut {NoStop}%
\bibitem [{\citenamefont {Carr}\ \emph {et~al.}(2018{\natexlab{b}})\citenamefont {Carr}, \citenamefont {Massatt}, \citenamefont {Torrisi}, \citenamefont {Cazeaux}, \citenamefont {Luskin},\ and\ \citenamefont {Kaxiras}}]{PhysRevB.98.224102}%
  \BibitemOpen
  \bibfield  {author} {\bibinfo {author} {\bibfnamefont {S.}~\bibnamefont {Carr}}, \bibinfo {author} {\bibfnamefont {D.}~\bibnamefont {Massatt}}, \bibinfo {author} {\bibfnamefont {S.~B.}\ \bibnamefont {Torrisi}}, \bibinfo {author} {\bibfnamefont {P.}~\bibnamefont {Cazeaux}}, \bibinfo {author} {\bibfnamefont {M.}~\bibnamefont {Luskin}},\ and\ \bibinfo {author} {\bibfnamefont {E.}~\bibnamefont {Kaxiras}},\ }\bibfield  {title} {\bibinfo {title} {Relaxation and domain formation in incommensurate two-dimensional heterostructures},\ }\href {https://doi.org/10.1103/PhysRevB.98.224102} {\bibfield  {journal} {\bibinfo  {journal} {Phys. Rev. B}\ }\textbf {\bibinfo {volume} {98}},\ \bibinfo {pages} {224102} (\bibinfo {year} {2018}{\natexlab{b}})}\BibitemShut {NoStop}%
\bibitem [{\citenamefont {Kang}\ and\ \citenamefont {Vafek}(2023)}]{PhysRevB.107.075408}%
  \BibitemOpen
  \bibfield  {author} {\bibinfo {author} {\bibfnamefont {J.}~\bibnamefont {Kang}}\ and\ \bibinfo {author} {\bibfnamefont {O.}~\bibnamefont {Vafek}},\ }\bibfield  {title} {\bibinfo {title} {Pseudomagnetic fields, particle-hole asymmetry, and microscopic effective continuum hamiltonians of twisted bilayer graphene},\ }\href {https://doi.org/10.1103/PhysRevB.107.075408} {\bibfield  {journal} {\bibinfo  {journal} {Phys. Rev. B}\ }\textbf {\bibinfo {volume} {107}},\ \bibinfo {pages} {075408} (\bibinfo {year} {2023})}\BibitemShut {NoStop}%
\bibitem [{\citenamefont {Tilak}\ \emph {et~al.}(2021)\citenamefont {Tilak}, \citenamefont {Lai}, \citenamefont {Wu}, \citenamefont {Zhang}, \citenamefont {Xu}, \citenamefont {Ribeiro}, \citenamefont {Canfield},\ and\ \citenamefont {Andrei}}]{Tilak2021}%
  \BibitemOpen
  \bibfield  {author} {\bibinfo {author} {\bibfnamefont {N.}~\bibnamefont {Tilak}}, \bibinfo {author} {\bibfnamefont {X.}~\bibnamefont {Lai}}, \bibinfo {author} {\bibfnamefont {S.}~\bibnamefont {Wu}}, \bibinfo {author} {\bibfnamefont {Z.}~\bibnamefont {Zhang}}, \bibinfo {author} {\bibfnamefont {M.}~\bibnamefont {Xu}}, \bibinfo {author} {\bibfnamefont {R.~d.~A.}\ \bibnamefont {Ribeiro}}, \bibinfo {author} {\bibfnamefont {P.~C.}\ \bibnamefont {Canfield}},\ and\ \bibinfo {author} {\bibfnamefont {E.~Y.}\ \bibnamefont {Andrei}},\ }\bibfield  {title} {\bibinfo {title} {Flat band carrier confinement in magic-angle twisted bilayer graphene},\ }\href {https://doi.org/10.1038/s41467-021-24480-3} {\bibfield  {journal} {\bibinfo  {journal} {Nature Communications}\ }\textbf {\bibinfo {volume} {12}},\ \bibinfo {pages} {4180} (\bibinfo {year} {2021})}\BibitemShut {NoStop}%
\bibitem [{\citenamefont {Yu}\ \emph {et~al.}(2023)\citenamefont {Yu}, \citenamefont {Wang}, \citenamefont {Katsnelson},\ and\ \citenamefont {Yuan}}]{PhysRevB.108.045138}%
  \BibitemOpen
  \bibfield  {author} {\bibinfo {author} {\bibfnamefont {G.}~\bibnamefont {Yu}}, \bibinfo {author} {\bibfnamefont {Y.}~\bibnamefont {Wang}}, \bibinfo {author} {\bibfnamefont {M.~I.}\ \bibnamefont {Katsnelson}},\ and\ \bibinfo {author} {\bibfnamefont {S.}~\bibnamefont {Yuan}},\ }\bibfield  {title} {\bibinfo {title} {Origin of the magic angle in twisted bilayer graphene from hybridization of valence and conduction bands},\ }\href {https://doi.org/10.1103/PhysRevB.108.045138} {\bibfield  {journal} {\bibinfo  {journal} {Phys. Rev. B}\ }\textbf {\bibinfo {volume} {108}},\ \bibinfo {pages} {045138} (\bibinfo {year} {2023})}\BibitemShut {NoStop}%
\bibitem [{\citenamefont {Tarnopolsky}\ \emph {et~al.}(2019)\citenamefont {Tarnopolsky}, \citenamefont {Kruchkov},\ and\ \citenamefont {Vishwanath}}]{PhysRevLett.122.106405}%
  \BibitemOpen
  \bibfield  {author} {\bibinfo {author} {\bibfnamefont {G.}~\bibnamefont {Tarnopolsky}}, \bibinfo {author} {\bibfnamefont {A.~J.}\ \bibnamefont {Kruchkov}},\ and\ \bibinfo {author} {\bibfnamefont {A.}~\bibnamefont {Vishwanath}},\ }\bibfield  {title} {\bibinfo {title} {Origin of magic angles in twisted bilayer graphene},\ }\href {https://doi.org/10.1103/PhysRevLett.122.106405} {\bibfield  {journal} {\bibinfo  {journal} {Phys. Rev. Lett.}\ }\textbf {\bibinfo {volume} {122}},\ \bibinfo {pages} {106405} (\bibinfo {year} {2019})}\BibitemShut {NoStop}%
\bibitem [{\citenamefont {Li}\ \emph {et~al.}(2022)\citenamefont {Li}, \citenamefont {Eaton}, \citenamefont {Fertig},\ and\ \citenamefont {Seradjeh}}]{PhysRevLett.128.026404}%
  \BibitemOpen
  \bibfield  {author} {\bibinfo {author} {\bibfnamefont {Y.}~\bibnamefont {Li}}, \bibinfo {author} {\bibfnamefont {A.}~\bibnamefont {Eaton}}, \bibinfo {author} {\bibfnamefont {H.~A.}\ \bibnamefont {Fertig}},\ and\ \bibinfo {author} {\bibfnamefont {B.}~\bibnamefont {Seradjeh}},\ }\bibfield  {title} {\bibinfo {title} {Dirac magic and lifshitz transitions in aa-stacked twisted multilayer graphene},\ }\href {https://doi.org/10.1103/PhysRevLett.128.026404} {\bibfield  {journal} {\bibinfo  {journal} {Phys. Rev. Lett.}\ }\textbf {\bibinfo {volume} {128}},\ \bibinfo {pages} {026404} (\bibinfo {year} {2022})}\BibitemShut {NoStop}%
\bibitem [{\citenamefont {Polizzi}(2009)}]{PhysRevB.79.115112}%
  \BibitemOpen
  \bibfield  {author} {\bibinfo {author} {\bibfnamefont {E.}~\bibnamefont {Polizzi}},\ }\bibfield  {title} {\bibinfo {title} {Density-matrix-based algorithm for solving eigenvalue problems},\ }\href {https://doi.org/10.1103/PhysRevB.79.115112} {\bibfield  {journal} {\bibinfo  {journal} {Phys. Rev. B}\ }\textbf {\bibinfo {volume} {79}},\ \bibinfo {pages} {115112} (\bibinfo {year} {2009})}\BibitemShut {NoStop}%
\bibitem [{\citenamefont {Lehmann}\ and\ \citenamefont {Taut}(1972)}]{pssb.2220540211}%
  \BibitemOpen
  \bibfield  {author} {\bibinfo {author} {\bibfnamefont {G.}~\bibnamefont {Lehmann}}\ and\ \bibinfo {author} {\bibfnamefont {M.}~\bibnamefont {Taut}},\ }\bibfield  {title} {\bibinfo {title} {On the numerical calculation of the density of states and related properties},\ }\href {https://doi.org/10.1002/pssb.2220540211} {\bibfield  {journal} {\bibinfo  {journal} {physica status solidi (b)}\ }\textbf {\bibinfo {volume} {54}},\ \bibinfo {pages} {469} (\bibinfo {year} {1972})}\BibitemShut {NoStop}%
\bibitem [{\citenamefont {Alon}\ and\ \citenamefont {Cederbaum}(2003)}]{alon2003}%
  \BibitemOpen
  \bibfield  {author} {\bibinfo {author} {\bibfnamefont {O.~E.}\ \bibnamefont {Alon}}\ and\ \bibinfo {author} {\bibfnamefont {L.~S.}\ \bibnamefont {Cederbaum}},\ }\bibfield  {title} {\bibinfo {title} {Hellmann-{{Feynman}} theorem at degeneracies},\ }\href {https://doi.org/10.1103/PhysRevB.68.033105} {\bibfield  {journal} {\bibinfo  {journal} {Phys. Rev. B}\ }\textbf {\bibinfo {volume} {68}},\ \bibinfo {pages} {033105} (\bibinfo {year} {2003})}\BibitemShut {NoStop}%
\bibitem [{\citenamefont {Kresse}\ and\ \citenamefont {Furthm{\"u}ller}(1996{\natexlab{a}})}]{kresse1996}%
  \BibitemOpen
  \bibfield  {author} {\bibinfo {author} {\bibfnamefont {G.}~\bibnamefont {Kresse}}\ and\ \bibinfo {author} {\bibfnamefont {J.}~\bibnamefont {Furthm{\"u}ller}},\ }\bibfield  {title} {\bibinfo {title} {Efficient iterative schemes for ab initio total-energy calculations using a plane-wave basis set},\ }\href {https://doi.org/10.1103/PhysRevB.54.11169} {\bibfield  {journal} {\bibinfo  {journal} {Phys. Rev. B}\ }\textbf {\bibinfo {volume} {54}},\ \bibinfo {pages} {11169} (\bibinfo {year} {1996}{\natexlab{a}})}\BibitemShut {NoStop}%
\bibitem [{\citenamefont {Kresse}\ and\ \citenamefont {Furthm{\"u}ller}(1996{\natexlab{b}})}]{kresse1996a}%
  \BibitemOpen
  \bibfield  {author} {\bibinfo {author} {\bibfnamefont {G.}~\bibnamefont {Kresse}}\ and\ \bibinfo {author} {\bibfnamefont {J.}~\bibnamefont {Furthm{\"u}ller}},\ }\bibfield  {title} {\bibinfo {title} {Efficiency of ab-initio total energy calculations for metals and semiconductors using a plane-wave basis set},\ }\href {https://doi.org/10.1016/0927-0256(96)00008-0} {\bibfield  {journal} {\bibinfo  {journal} {Computational Materials Science}\ }\textbf {\bibinfo {volume} {6}},\ \bibinfo {pages} {15} (\bibinfo {year} {1996}{\natexlab{b}})}\BibitemShut {NoStop}%
\bibitem [{\citenamefont {Bl{\"o}chl}(1994)}]{blochl1994}%
  \BibitemOpen
  \bibfield  {author} {\bibinfo {author} {\bibfnamefont {P.~E.}\ \bibnamefont {Bl{\"o}chl}},\ }\bibfield  {title} {\bibinfo {title} {Projector augmented-wave method},\ }\href {https://doi.org/10.1103/PhysRevB.50.17953} {\bibfield  {journal} {\bibinfo  {journal} {Phys. Rev. B}\ }\textbf {\bibinfo {volume} {50}},\ \bibinfo {pages} {17953} (\bibinfo {year} {1994})}\BibitemShut {NoStop}%
\bibitem [{\citenamefont {Kresse}\ and\ \citenamefont {Joubert}(1999)}]{kresse1999}%
  \BibitemOpen
  \bibfield  {author} {\bibinfo {author} {\bibfnamefont {G.}~\bibnamefont {Kresse}}\ and\ \bibinfo {author} {\bibfnamefont {D.}~\bibnamefont {Joubert}},\ }\bibfield  {title} {\bibinfo {title} {From ultrasoft pseudopotentials to the projector augmented-wave method},\ }\href {https://doi.org/10.1103/PhysRevB.59.1758} {\bibfield  {journal} {\bibinfo  {journal} {Phys. Rev. B}\ }\textbf {\bibinfo {volume} {59}},\ \bibinfo {pages} {1758} (\bibinfo {year} {1999})}\BibitemShut {NoStop}%
\bibitem [{\citenamefont {Klime{\v s}}\ \emph {et~al.}(2011)\citenamefont {Klime{\v s}}, \citenamefont {Bowler},\ and\ \citenamefont {Michaelides}}]{klimes2011}%
  \BibitemOpen
  \bibfield  {author} {\bibinfo {author} {\bibfnamefont {J.}~\bibnamefont {Klime{\v s}}}, \bibinfo {author} {\bibfnamefont {D.~R.}\ \bibnamefont {Bowler}},\ and\ \bibinfo {author} {\bibfnamefont {A.}~\bibnamefont {Michaelides}},\ }\bibfield  {title} {\bibinfo {title} {Van der {{Waals}} density functionals applied to solids},\ }\href {https://doi.org/10.1103/PhysRevB.83.195131} {\bibfield  {journal} {\bibinfo  {journal} {Phys. Rev. B}\ }\textbf {\bibinfo {volume} {83}},\ \bibinfo {pages} {195131} (\bibinfo {year} {2011})}\BibitemShut {NoStop}%
\bibitem [{\citenamefont {{Rom{\'a}n-P{\'e}rez}}\ and\ \citenamefont {Soler}(2009)}]{roman-perez2009}%
  \BibitemOpen
  \bibfield  {author} {\bibinfo {author} {\bibfnamefont {G.}~\bibnamefont {{Rom{\'a}n-P{\'e}rez}}}\ and\ \bibinfo {author} {\bibfnamefont {J.~M.}\ \bibnamefont {Soler}},\ }\bibfield  {title} {\bibinfo {title} {Efficient {{Implementation}} of a van der {{Waals Density Functional}}: {{Application}} to {{Double-Wall Carbon Nanotubes}}},\ }\href {https://doi.org/10.1103/PhysRevLett.103.096102} {\bibfield  {journal} {\bibinfo  {journal} {Phys. Rev. Lett.}\ }\textbf {\bibinfo {volume} {103}},\ \bibinfo {pages} {096102} (\bibinfo {year} {2009})}\BibitemShut {NoStop}%
\end{thebibliography}%

\end{document}